\newcommand{\AP}[3]{Ann.\ Phys.\ {\bf #1},\ #2 (#3)}
\newcommand{\NPA}[3]{Nucl.\ Phys.\ {\bf A#1},\ #2 (#3)}
\newcommand{\NPB}[3]{Nucl.\ Phys.\ {\bf B#1},\ #2 (#3)}

\newcommand{\PLB}[3]{Phys.\ Lett.\ B\ {\bf #1},\ #2 (#3)}
\newcommand{\PR}[3]{Phys.\ Rep.\ {\bf #1},\ #2 (#3)}
\newcommand{\PRL}[3]{Phys.\ Rev.\ Lett.\ {\bf #1},\ #2 (#3)}
\newcommand{\PRA}[3]{Phys.\ Rev.\ A\ {\bf #1},\ #2 (#3)}

\newcommand{\PRC}[3]{Phys.\ Rev.\ C\ {\bf #1},\ #2 (#3)}
\newcommand{\PRD}[3]{Phys.\ Rev.\ D\ {\bf #1},\ #2 (#3)}
\newcommand{\JPG}[3]{J.\ Phys.\ G\ {\bf #1},\ #2 (#3)}

\newcommand{\ZPC}[3]{Z.\ Phys.\ C\ {\bf #1},\ #2 (#3)}

\newcommand{\EPJA}[3]{Eur.\ Phys.\ J.\ A\ {\bf #1},\ #2 (#3)}
\newcommand{\PTP}[3]{Prog.\ Theo.\ Phys.\ {\bf #1},\ #2 (#3)}

\renewcommand\k{\kappa}

\newcommand{\diracslash}[1]{#1\llap{/\kern2pt}}

\newcommand{\be}{\begin{equation}}
\newcommand{\ee}{\end{equation}}
\newcommand{\bea}{\begin{eqnarray}}
\newcommand{\eea}{\end{eqnarray}}
\newcommand{\ba}[1]{\begin{array}{#1}}
\newcommand{\ea}{\end{array}}

\documentclass[prd,aps,floats,nofootinbib,tightenlines,showpacs,twocolumn,floatfix]{revtex4}
\usepackage{graphicx}
\begin{document}
\title { BCS BEC crossover and phase structure of relativistic system:
a variational approach}
\author{Bhaswar Chatterjee}
\email{bhaswar@prl.res.in}
\affiliation{Theory Division, Physical Research Laboratory,
Navrangpura, Ahmedabad 380 009, India}
\author{Hiranmaya Mishra}
\affiliation{Theory Division, Physical Research Laboratory,
Navrangpura, Ahmedabad 380 009, India}
\affiliation{School of Physical Sciences, Jawaharlal Nehru University, New Delhi-110067,
India}
\email{hm@prl.res.in}
\author{Amruta Mishra}
\affiliation{Department of Physics, Indian Institute of Technology, New 
Delhi-110016,India}
\email{amruta@physics.iitd.ac.in}
\affiliation{Frankfurt Institute for Advanced Studies,
Universit\"at Frankfurt, D-60438 Frankfurt, Germany}

\date{\today} 

\def\be{\begin{equation}}
\def\ee{\end{equation}}
\def\bearr{\begin{eqnarray}}
\def\eearr{\end{eqnarray}}
\def\zbf#1{{\bf {#1}}}
\def\bfm#1{\mbox{\boldmath $#1$}}
\def\hf{\frac{1}{2}}
\begin{abstract}
We investigate here the BCS BEC crossover in relativistic systems 
using a variational construct for the ground state and the minimization
of the thermodynamic potential.
This is first studied in a four fermion point interaction model and with a 
BCS type ansatz for the ground state with fermion pairs. It is shown
that the antiparticle degrees of freedom play an important role in 
the BCS BEC crossover physics, even when the ratio of fermi momentum 
to the mass of the fermion is small.
We also consider the phase structure for the case of fermion pairing 
with imbalanced populations.
Within the ansatz, thermodynamically stable gapless modes for 
both fermions and anti fermions are seen for strong coupling in the BEC regime.
We further investigate the effect of fluctuations of the condensate 
field by treating it as a dynamical field and generalize the BCS 
ansatz to include quanta of the condensate field also in a boson fermion
model with quartic self interaction of the condensate field. It is seen 
that the critical temperature decreases with inclusion of fluctuations.  
 
\end{abstract}

\pacs{12.38.Mh, 24.85.+p} 

\maketitle

 \section{Introduction}

Color superconductivity has become an active field of research during 
the last few years in the field of strong interaction physics \cite{review}.
At asymptotically high density the ground state of QCD is shown to be a 
color superconductor from first principle calculations \cite{pisarski}. However,
for intermediate densities, relevant for the matter in the core of 
neutron stars, the perturbation
theory breaks down and one uses effective models of strong interactions whose 
parameters are fitted to reproduce low energy hadronic properties. Although,
it is reassuring that certain quantities like the superconducting gap
estimated from weak coupling QCD or using effective models like 
Nambu Jona Lasinio (NJL) model yield similar magnitudes, the strong 
coupling dynamics at not so high densities remain uncontrolled. 
As the density decreases, before the quarks are confined,
the coupling can become large enough so that the coherence length
in the superconducting phase can be
of the order of inter particle 
separation \cite{hatsuda,rischke}.  When this happens it is natural 
to imagine the 
quark quark Cooper pairs as localized bound states rather than
describing  them as an extended macroscopic medium. 
At low enough temperature
the ground state can then become a Bose Einstein condensate (BEC) 
of diquark molecules. At still lower densities this diquark matter can undergo
a confinement transition to form hadronic matter. The scenario that we have
in mind is similar to as in Ref. \cite{pirner}, where, as the baryon density 
is increased, the baryonic matter undergoes a phase transition to diquark 
BEC phase.
 The transition from BCS to BEC is most likely to be a 
 crossover similar to their non relativistic counterparts in 
 condensed matter systems like cold fermionic atoms.
Various methods have been used recently to describe relativistic BCS - BEC crossover
within a four fermion point interaction model \cite{zhuang,andreas,rischke}. 
Effects of fluctuations have also been studied in the recent past to go 
beyond the mean field
approximations \cite{zhuangb,abukib,abuki,deng} using a gaussian approximation
as well as including collective excitations \cite{brauner}. 
Another interesting feature for the matter in the interior of
neutron star is the variety of exotic non BCS 
phases that arise when kinematical constraints like neutrality with
respect to color and 
electric charges are imposed. Such kind of stressed pairing has attracted
attention  both in quark matter \cite{krisandreas} and in polarized 
Fermi gas of atoms
\cite{nardulli,amhmloff}.
 Relativistic BCS-BEC crossover was also studied in a boson fermion model and 
a rich phase structure was observed in such a system when there is a mismatch in
 the number densities of the condensing fermions \cite{andreas}.

Our approach to such problems in quark matter \cite {hmparikh,amhma,amhmb} 
as well as in cold atoms \cite{rapid} has been variational. Charge neutral 
matter was considered with an explicit construct for the ground state in 
terms of
quark pair operators as well as quark antiquark operators. The ansatz functions 
in this construct were determined from the minimization of the thermodynamic
potential subject to the constraints of neutrality conditions with respect to
color and electrical charges. The effect of six quark determinant 
interaction was also considered which highlighted the consequences of
strange quark mass in such stressed pairing cases \cite{amhmc}. We apply here
similar method to study BCS-BEC crossover (as well as possible phase transition
for mismatch in number densities) for a relativistic system in a 
fermionic theory with a four fermi point interaction model. As we shall 
demonstrate here the simple BCS type of ansatz for the ground state 
leads to results similar to the mean field results of
a relativistic Bose Fermi model \cite{andreas}. This is demonstrated by 
taking a BCS type ansatz for the ground state and treating the condensate field 
as a classical auxiliary field. We next treat the condensate scalar field as a 
dynamical field and generalize the BCS ansatz to include the quanta of this
condensate fields in a Bose Fermi model with a quartic self interaction for the
scalar field. This improved ansatz for the ground state leads to a
mass gap equation for the scalar field and the corresponding thermodynamic 
potential that can be obtained by resummation of bubble diagrams of
perturbation theory similar to the results obtained in
Cornwall-Jackiew Tombolis (CJT) composite operator formalism
\cite{cameliapi,rislena}.

We organize the paper as follows. In section II we shall consider a relativistic model
with a four Fermi point interaction to discuss the BCS-BEC crossover
physics. We shall take the bosonic condensate field as a classical auxiliary
field while retaining the quantum nature of the fermionic field. In subsection 
II A, we spell out the ansatz for the ground state and in subsection II B, we 
evaluate the thermodynamic potential and consider the minimization of the same
with respect to the ansatz functions. In section III, we solve the gap
equation and the number density equations and evaluate the thermodynamic
potential. The thermodynamic potential for different phases are compared
in this section to decide which phase is thermodynamically stable at what
coupling and at what difference in the chemical potential. We also discuss our 
numerical results in this section. The results obtained in this section
using the variational ansatz,
is similar to that obtained within mean field approximations in earlier calculations regarding BCS -BEC crossover 
and phase structure for imbalanced populations \cite{zhuang,andreas,rischke}. In section IV, we treat the
scalar condensate field as a dynamical field and generalise the ansatz 
of section III to include the quanta of the
fluctuations over and above the BCS ground state. We investigate this in a model with quartic self interaction
for the scalar field, calculating the mass of the scalar field self consistently. We also include the effect of temperature
in these calculations.  Finally we summarize our results in section \ref{summary}.

\section{A fermionic  model for relativistic superfluid}
\label{model}

We shall consider a general relativistic model with only fermionic
degrees of freedom in the Lagrangian. In particular we consider a general form
for the Lagrangian given as
\be
{\cal L}= \bar\psi^i(i\gamma^\mu\partial_\mu-m+\mu_i\gamma^0)\psi^i
+{\cal L}_I\equiv{\cal L}_f +{\cal L}_I
\label{lag}
\ee
where, $\psi^i$,$\bar\psi^i$ denote the Dirac fields with indices $i=1,2$
denoting the 'flavor' for the fermions. Further, 
we have taken, for simplicity, both the fermions with the 
same mass $m$. With different masses, the fermi energies or 
the chemical potential in the weak coupling, can be different 
for the same number densities of the two pairing species.
In the present analysis, for simplicity, we assume the masses of the 
pairing fermions to be the same, but consider their chemical potentials to be
different.

To study superfluidity that results from Cooper pairing of two different 
flavors of fermions, we introduce the following local interaction term in the
Lagrangian,
\be
{\cal L}_I=-G(\bar\psi_c^i\gamma^5\psi^j|\epsilon^{ij}|)(\bar\psi^k\gamma^5
\psi_c^l|\epsilon^{kl}|)
\label{li}
\ee
Here, $G$ is the coupling constant, $\psi_c=C\bar\psi^T$, $\bar \psi_c=\psi^TC$,
with the charge conjugation matrix being $C=i\gamma^2\gamma^0$. The 
$|\epsilon^{ij}|$ term ensures the cross flavor, spin zero antisymmetric
pairing.

To illustrate the mean field calculation, introducing a field $\Phi$,
we can rewrite the interaction term as 
\be
{\cal L}_I=g|\epsilon^{ij}|\left(\bar\psi^i\gamma^5\psi^j_c\Phi+
\bar\psi^i_c\gamma^5\psi^j\Phi^* \right)-m_b^2\Phi^*\Phi
\label{liphi}
\ee
with $G=g^2/m_b^2$ which can be identified upon elimination of the field $\Phi$.
We note that in the absence of a kinetic term the field $\Phi$ is an auxiliary
field whose value in terms of the fermion bilinears is known. In the following,
however, we shall treat the field $\Phi$ as a classical field and replace it
by its expectation value while retaining the quantum nature for the fermion
field. This will enable us to calculate the effective potential as 
a function of $\phi_0=\langle\Phi\rangle$.

\subsection{ The ansatz for the ground state}

To make the notations clear, let us  first note that the fermion field operator
expansion in momentum space is given as
 \cite{hmnj,amspm}
\begin{eqnarray}
&&\psi (\zbf x )=\frac{1}{(2\pi)^{3/2}}\int \tilde\psi(\zbf k)
e^{i\zbf k\cdot\zbf x}d\zbf k \nonumber\\ 
&=&\frac{1}{(2\pi)^{3/2}}\int \left[U_0(\zbf k)q(\zbf k )
+V_0(-\zbf k)\tilde q(-\zbf k )\right]e^{i\zbf k\cdot \zbf x}d \zbf k
\label{psiexp} \;\;\;
\end{eqnarray}
where,
\begin{eqnarray}
U_0(\zbf k )=&&\left(\begin{array}{c}\cos(\frac{\chi^0}{2})\\
\zbf \sigma \cdot \hat k \sin(\frac{\chi^0}{2})
\end{array}\right),\;\;
\nonumber\\
V_0(-\zbf k )=&&
\left( \begin{array}{c} -\zbf \sigma \cdot \hat k  \sin(\frac{\chi^0}{2})
\\ \cos(\frac{\chi^0}{2})\end{array}
\right).
\label{uv0}
\end{eqnarray}
The operators $q$ and $\tilde q$ are the
two component particle annihilation and antiparticle creation operators 
respectively which annihilate or create quanta acting upon
the perturbative vacuum $|0\rangle$. We have suppressed 
here the flavor indices of the fermion field operators. The function 
$\chi^0(\zbf k)$ in the spinors in Eq.(\ref{uv0}) are given as 
$\cot{\chi_i^0}=m_i/|\zbf k|$,  for free massive fermion fields, 
$i$ being the flavor index. For massless fields
$\chi^0(|\zbf k|)=\pi/2$. We shall in the following, however consider
for simplicity, $m_1=m_2=m$. 
We shall use the notations and conventions of Ref. \cite{amhma,amhmb} and
recapitulate
briefly the construction of the variational ansatz for the ground state.
We take  it  as a squeezed
coherent state fermion condensates given as
\cite{amhma,amhmb,hmparikh} 
\begin{equation} 
|\Omega\rangle= {\cal U}_d|0\rangle.
\label{u0}
\end{equation} 

Here, ${\cal U}_d$ is a unitary operator which creates
fermion pairs. Explicitly,
the operator, ${\cal U}_d$ is given as
\begin{equation}
{\cal U}_d=\exp(B_d^\dagger-B_d),
\label{omg}
\end{equation}
where, $B_d^\dagger$ is the pair creation operator as given by
\begin{eqnarray}
{B}_d ^\dagger&=&\int \left[q_r^{i}(\zbf k)^\dagger
r f(\zbf k) q_{-r}^{j}(-\zbf k)^\dagger
|\epsilon_{ij}|\right]d\zbf k \nonumber\\
&+&\int \left[\tilde q_r^{i}(\zbf k)
r f_1(\zbf k) \tilde q_{-r}^{j}(-\zbf k)
|\epsilon_{ij}|\right]
d\zbf k.
\label{bd}
\end{eqnarray}
\noindent 
In the above, $i,j$ are flavor indices,
and $r(=\pm 1/2) $ is the spin index.
Further, $f(\zbf k)$ and $f_1(\zbf k)$ are ansatz functions
associated with fermion pairs and anti fermion pairs describing the condensates
and shall be determined by the minimization of the thermodynamic potential.
Note that we have assumed these `condensate functions' $f(\zbf k)$
 and $f_1(\zbf k)$ to be
independent of flavor  indices. We give a post-facto justification 
for this to be that the functions depend upon the {\it{ average}} energy 
and {\it average}
chemical potential of the fermions/anti-fermions that condense.

Finally, to include the effects of temperature and density we  write
down the state at finite temperature and density 
$|\Omega(\beta,\mu)\rangle$  taking
a thermal Bogoliubov transformation over the state $|\Omega\rangle$ 
using thermo field dynamics (TFD) as described in Ref.s \cite{tfd,amph4}.
We then have,
\begin{equation} 
|\Omega(\beta,\mu)\rangle={\cal U}_{\beta,\mu}|\Omega\rangle={\cal U}_{\beta,\mu}
{\cal U}_d |0\rangle,
\label{ubt}
\end{equation} 
where ${\cal U}_{\beta,\mu}$ is given as
\begin{equation}
{\cal U}_{\beta,\mu}=e^{{\cal B}^{\dagger}(\beta,\mu)-{\cal B}(\beta,\mu)},
\label{ubm}
\end{equation}
with 
\begin{eqnarray}
{\cal B}^\dagger(\beta,\mu)&=&\int \Big [
q^\prime (\zbf k)^\dagger \theta_-(\zbf k, \beta,\mu)
\underline q^{\prime} (\zbf k)^\dagger\big ] d\zbf k \nonumber\\
&+&\int \big [\tilde q^\prime (\zbf k) \theta_+(\zbf k, \beta,\mu)
\underline { \tilde q}^{\prime} (\zbf k)\Big ] d\zbf k.
\label{bth}
\end{eqnarray}
In Eq.(\ref{bth}) the ansatz functions $\theta_{\pm}(\zbf k,\beta,\mu)$
will be related to quark and antiquary distributions, and, the underlined
operators are the operators in the extended Hilbert space associated with
thermal doubling in TFD method. In Eq.(\ref{bth}) we have suppressed
the flavor indices in the fermion operators as well as in the functions
$\theta(\zbf k,\beta,\mu)$.

 All the functions in the ansatz in Eq.(\ref{ubt})
are to be obtained by minimizing the
thermodynamic potential.
We shall carry out this minimization
in the next subsection.

\subsection{Evaluation of thermodynamic potential and gap equations }
\label{evaluation}
To calculate the thermodynamic potential corresponding to the Lagrangian
of Eq.(\ref{lag}) and Eq.(\ref{liphi}) and the state given in Eq.(\ref{ubt}),
we first write down the expectation values of the following fermion operator
bilinears.

\begin{eqnarray}
\langle \Omega(\beta,\mu)
 |\tilde\psi_\gamma^{i}(\zbf k)\tilde\psi ^{j}_\delta(\zbf k')^{\dagger}
|\Omega(\beta,\mu)\rangle \nonumber\\
=\delta^{ij}
\Lambda_{+\gamma\delta}^{i}(\zbf k,\beta,\mu)\delta(\zbf k-\zbf k')
\label{psipsidb}
\end{eqnarray}
and,
\begin{eqnarray}
\langle \Omega(\beta,\mu)
|\tilde\psi_\delta^{i\dagger}(\zbf k)\tilde\psi_\gamma^{j}(\zbf k')
|\Omega(\beta,\mu)\rangle \nonumber\\
=\delta^{ij}
\Lambda_{-\gamma\delta}^{i}(\zbf k,\beta,\mu)\delta(\zbf k-\zbf k')
\label{psidpsib}
\end{eqnarray}
where,
\begin{eqnarray}
&&\Lambda_{\pm \gamma\delta}^{i}(\zbf k,\beta,\mu)
 =  \hf\left[1\pm \left( F_1^{i}(\zbf k)
-F^{i}(\zbf k)\right)\pm \big(\gamma^0\cos\chi^i (\zbf k)\right . \nonumber\\
& + & \left . \bfm\alpha\cdot\hat\zbf k\sin \chi^i
(\zbf k)\big)\big(1-F^{i}(\zbf k)-F_1^{i}(\zbf k)\big)
\right]_{\gamma\delta}.
\label{prpb}
\end{eqnarray}
In the above, 
$\tilde\psi(\zbf k)$ is the Fourier transform of $\psi(\zbf x)$ \cite{amhmb}.
The effect of the fermion condensates and their temperature
 and/or density
dependences are encoded in the functions $F^{i}(\zbf k)$ and $F_1^{i}
(\zbf k)$ given as
\begin{equation}
F^{i}(\zbf k)=\left(\sin^2\theta^{i}_-(\zbf k)+\sin^2 f(\zbf k)
\cos 2\theta_-^{i,j}(\zbf k)
\right)
\label{fkb}
\end{equation}
and,
\begin{equation}
F_1^{i}(\zbf k)=\left(\sin^2\theta^{i}_+(\zbf k)+\sin^2 f_1(\zbf k)
\cos 2\theta_+^{i,j}(\zbf k)
\right).
\label{f1kb}
\end{equation}
We have defined $\cos 2\theta_{\pm}^{i,j}=1-\sin^2\theta^{i}_{\pm}-
\sin^2\theta_{\pm}^{j}$ with $i\neq j$.

\noindent
For difermion operators, we have,

\bearr
\langle \Omega(\beta,\mu)| \psi^{i}_\alpha(\zbf x)\psi^{j}_\gamma
(\zbf 0)
|\Omega(\beta,\mu) \rangle \nonumber\\
=-\frac{1}{(2\pi)^3}
\int e^{i\zbf k\cdot\zbf x}
{\cal {P}}_{+\gamma\alpha}^{i,j}(\zbf k,\beta,\mu)d\zbf k
\eearr
and,
\bearr
\langle \Omega(\beta,\mu)| \psi^{i\dagger}_\alpha(\zbf x)
\psi^{j\dagger}_\gamma
(\zbf 0)
|\Omega(\beta,\mu) \rangle \nonumber\\
=-\frac{1}{(2\pi)^3}\int e^{i\zbf k\cdot\zbf x}
{\cal {P}}_{-\alpha\gamma}^{i,j}(\zbf k,\beta,\mu)d\zbf k,
\label{psi}
\eearr

where,
\begin{widetext} 
\bearr
{\cal{P}}_+^{i,j}(\zbf k,\beta,\mu)
&=&\frac{|\epsilon^{ij}|}{4}\bigg[S^{i,j}(\zbf k)\left(
\cos\left(\frac{\chi_i-\chi_j}{2}\right)
-\bfm\gamma\cdot\hat\zbf k\sin\left(\frac{\chi_i-\chi_j}{2}\right)\right)
\nonumber\\
&+&\left(\gamma^0
\cos \left(\frac{\chi_{i}+\chi_j}{2}\right)-\bfm\alpha\cdot
\hat\zbf k\sin\left(\frac{\chi_i+\chi_j}{2}\right)\right)A^{i,j}(\zbf k)
\bigg]\gamma_5 C \\
\label{calpp}
{\cal{P}}_-^{i,j}(\zbf k,\beta,\mu)
&=&\frac{|\epsilon^{ij}|
C\gamma_5}{4}\bigg [S^{i,j}(\zbf k)\left(
\cos\left(\frac{\chi_i-\chi_j}{2}\right)
+\bfm\gamma\cdot\hat\zbf k\sin\left(\frac{\chi_i-\chi_j}{2}\right)\right)
\nonumber\\
&+&
\left(\gamma^0\cos\left(\frac{\chi_i+\chi_j}{2}\right)
-\bfm\alpha\cdot \hat\zbf k\sin \left(\frac{\chi_i+\chi_j}{2}\right)
\right)A^{i,j}(\zbf k)
\bigg]
\label{calpm}
\eearr
\end{widetext}

\noindent Here, $C=i\gamma^2 \gamma^0$ is the charge conjugation matrix (we
use the notation of Bjorken and Drell) and the functions $S(\zbf k)$ and 
$A(\zbf k)$ are given as,
\begin{eqnarray}
S^{i,j}(\zbf k)&=&\sin\!2f(\zbf k) \cos 2\theta^{i,j}_-(\zbf k,\beta,\mu) \nonumber\\
&+&\sin\!2f_1(\zbf k)\cos 2\theta^{i,j}_+(\zbf k,\beta,\mu),
\label{sk}
\end{eqnarray}

and,
\begin{eqnarray}
A^{i,j}(\zbf k)&=&\sin\!2f(\zbf k)\cos 2\theta^{i,j}_-(\zbf k,\beta,\mu) \nonumber\\
&-&\sin\!2f_1(\zbf k)\cos 2\theta_+^{i,j}(\zbf k,\beta,\mu),
\label{ak}
\end{eqnarray}
These expressions are used to calculate thermal 
expectation value of the Hamiltonian and compute the thermodynamic
potential given as 
\be
\Omega=
\epsilon-\mu^i\rho^i-\frac{1}{\beta}s
\label{omgx}
\ee
where $\epsilon$ is the energy density and $s$ is the entropy density
and $\rho^i=\langle\psi^{i\dagger}\psi^i\rangle$ $(i=1,2)$ is the 
number density of $i$-th species. 

It is then straightforward  to calculate the expectation value of the 
Hamiltonian corresponding to the Lagrangian given in Eq.(\ref{lag}) and
Eq.(\ref{liphi}).
This can be written as
\be
\epsilon -\mu^i\rho^i= \langle H-\mu^i \psi^{i\dagger}\psi^i \rangle = T  +V_D 
\ee
 Explicitly,
the kinetic energy  minus the $\mu^i\rho^i$ part is given as 
\begin{widetext}
\bearr
T \equiv 
 \langle \Omega(\beta,\mu)|
\psi_i^\dagger
(-i\bfm \alpha \cdot \bfm \nabla +\gamma^0 m-\mu^i )\psi_i
| \Omega(\beta,\mu)\rangle
 =  \frac{2}{(2\pi)^3}\sum_{i=1}^2
\int d \zbf k \left( \sqrt{\zbf k^2+m^2}(F^{i}+F_1^{i})
-\mu^i(F^i-F_1^i)\right)
\label{tren}
\eearr
\end{widetext}
where, $F^{i}$ and $F_1^{i}$ are given by equations (\ref{fkb}) 
and (\ref{f1kb}). Here we have subtracted out the vacuum contributions.

Similarly, the contribution from the  interaction
from Eq.(\ref{liphi}) to the energy density is given as
\be
V_D=
-  \langle \Omega(\beta,\mu)|{\cal L}_I
| \Omega(\beta,\mu)\rangle
=-4g I_D\phi_0+m_b^2\phi_0^2
\label{vd}
\ee
where we have taken $\phi_0$ to be real. In the above,
\begin{widetext}
\bearr
I_D=\frac{1}{2}\langle\bar{\psi_c}^{i}\gamma^5|\epsilon^{ij}|\psi^{j}
\rangle =\frac{1}{(2\pi)^3}\int d{\zbf k}\left[
\sin 2f(\zbf k) (1-\sin^2\theta_-^1-\sin^2\theta_-^2)
+\sin 2f_1(\zbf k) (1-\sin^2\theta_+^1-
\sin^2\theta_+^2)\right]
\label{id}
\eearr
\end{widetext}
which is proportional to the fermion condensate.

Finally, to calculate the thermodynamic potential we have to include
the entropy density for the fermions. This is given as
\cite{tfd}
\begin{widetext}
\bearr
s = -\frac{2}{(2\pi)^3}\sum_{i}\int d \zbf k
\Big ( \sin^2\theta^{i}_-\ln \sin^2\theta^{i}_-
+\cos^2\theta^{i}_-\ln \cos^2\theta^{i}_- 
 + \sin^2\theta^{i}_+\ln \sin^2\theta^{i}_+
+\cos^2\theta^{i}_+\ln \cos^2\theta^{i}_+\Big ).
\label{ent}
\eearr
\end{widetext}
The extremisation of the thermodynamic potential Eq. (\ref{omgx}) with 
respect to the condensate functions $f(\zbf k)$ and $f_1(\zbf k)$
yields
\be
\tan2 f(\zbf k)=
\frac{2g\phi_0}{\bar\epsilon-\bar\mu}\equiv\frac{\Delta}{\bar\epsilon-\bar\mu}
\label{tan2f}
\ee

and
\be
\tan 2f_1(\zbf k)=
\frac{2g\phi_0}{\bar\epsilon+\bar\mu}\equiv\frac{\Delta}{\bar\epsilon+\bar\mu}
\label{tan2f1}
\ee
where, we have defined the superconducting gap 
$\Delta= 2g\phi_0$.
In the above $\bar\epsilon=(\epsilon_1+\epsilon_2)/2$ and 
$\bar\mu= (\mu_{1}+\mu_{2})/2$ with $\epsilon_i=\sqrt{\zbf k^2+m_i^2}$.
It is thus seen that the condensate functions depend upon
the {\em average} energy and the {\em average} chemical potential
of the fermions/anti-fermions that condense.

Finally, the minimization of the thermodynamic potential with respect to the
thermal functions $\theta_{\pm}(\zbf k)$ gives
\be
\sin^2\theta_\pm^{i}(\zbf k)=\frac{1}{\exp(\beta\omega_\pm^{(i)})+1}.
\label{them}
\ee
where,
\begin{mathletters}
\be
\omega_\pm^{(1)} =
\bar\omega_\pm +\delta_\epsilon\pm \delta_\mu
\label{omgpm1}
\ee
and
\be
\omega_\pm^{(2)} = 
\bar\omega_\pm -\delta_\epsilon\mp \delta_\mu
\label{omgpm2}
\ee
\end{mathletters}
\noindent Here, $\bar\omega_\pm=\sqrt{\Delta^2+\bar\xi_\pm^2}$, $\bar\xi_\pm
=(\xi_{1\pm}+\xi_{2\pm})/2$,
 $\xi_{i\pm}=\epsilon_i\pm\mu_i$. Further, the chemical potential 
 difference $\delta_\mu=
(\mu_1-\mu_2)/2$ and $\delta_\epsilon=
(\epsilon_1-\epsilon_2)/2$. Notice that for $\delta_\mu>0$ and for equal
masses for the two species $\delta_\epsilon=0$ with $\epsilon_1
=\epsilon_2=
\epsilon=\sqrt{\zbf k^2+m^2}$
, we can have the  possibility of
gapless modes for $\omega_-^{(1)}$ or $\omega_+^{(2)}$.

Using these dispersion relations, and 
substituting the condensate functions and the distribution functions,
leads to the thermodynamics potential given by equation (\ref{omgx}), as
\bearr
\Omega&=&
\frac{2}{(2\pi)^3}\int\left(2\epsilon-\bar\omega_--\bar\omega_+\right)
d\zbf k\nonumber\\
&-&
\frac{2}{(2\pi)^3\beta}\sum_i\int d\zbf k \bigg[\ln\left(1+e^{(-\beta\omega_-^{(i)})}\right)
\nonumber\\
&+&\ln\left(1+e^{(-\beta\omega_+^{(i)})}\right)\bigg] +m_b^2\phi_0^2.
\label{thpot}
\eearr
Here the extremisation over $\phi_0$ is yet to be done. An extremisation
with respect to $\phi_0$ leads to the gap equation
\be
\frac{m_b^2}{4g^2}=\int\frac{d\zbf k}{(2\pi)^3}\left[
\frac{\cos 2\theta_-^{1,2}}{\bar\omega_-}+\frac{\cos 2\theta_+^{1,2}}{\bar\omega_+}
\right],
\label{gapeq}
\ee
with $\cos 2\theta_\mp^{1,2}=1-\sin^2\theta^1_\mp-\sin^2\theta^2_\mp$.


The gap equation is quadratically divergent which is rendered finite
in the NJL model with a momentum cut off $\Lambda$. In the non relativistic
case this is rendered finite by subtracting out  the vacuum contribution
and relating the four fermion coupling to
the s-wave scattering length \cite{randeria,rapid}. A similar approach
can be done for the relativistic case also by relating the coupling to the 
s-wave scattering length. This leads to the renormalized gap equation
\cite{zhuang}
\bearr
-\frac{m}{4\pi a}&=&\int\frac{d\zbf k}{(2\pi)^3}\bigg[
\frac{\cos 2\theta_-^{1,2}}{\bar\omega_-}+\frac{\cos 2\theta_+^{1,2}}{\bar\omega_+}\nonumber\\
&-&\frac{1}{\epsilon-m}-\frac{1}{\epsilon+m}
\bigg].
\label{gapr}
\eearr
However, after this subtraction, unlike the nonrelativistic case, the ultraviolet
cutoff dependence is still there in the above gap equation,
although the dependence is milder.
We might also note that one could have defined a renormalized boson mass
$m_{b,r}$ with
$m_{b,r}^2=\partial\Omega/\partial\phi_0^2|_{\phi_0=T=0,\mu=m}$
as in Ref.\cite{andreas} and one would have arrived at the same gap equation.
For the present, we shall take the renormalized coupling as in
Eq.(\ref{gapr}) and treat this as the crossover parameter. As a function 
of this coupling one has to calculate the gap parameter $\Delta$ 
for different densities of the fermions of the two species. 
The average number density is  given as
\be
\bar\rho=\frac{\rho_1+\rho_2}{2}=\rho_--\rho_+,
\label{rhob}
\ee
 where, the fermionic component is given as
 \be
 \rho_-=
-\frac{1}{(2\pi)^3}\int
\frac{\xi_-}{\bar\omega_-}\cos 2\theta_-^{1,2}d\zbf k
\label{rhom}
\ee
and the anti fermionic component of the average number density is
\be
\rho_+=-\frac{1}{(2\pi)^3}\int
\frac{\xi_+}{\bar\omega_+}\cos 2\theta_+^{1,2}
d\zbf k
\label{rhop}
\ee
where, $\cos 2\theta_{\mp}^{1,2}=(1-\sin^2\theta_{\mp}^1(\zbf k)
-\sin^2\theta_{\mp}^2(\zbf k))$
with  $\sin^2\theta_\pm^i(\zbf k)$ being the thermal distribution functions
for the fermions defined in Eq.(\ref{them}).
The difference in the number densities is given as
\begin{eqnarray}
\delta_\rho=\frac{\rho_1-\rho_2}{2} & = & \frac{1}{(2\pi)^3}\int
\left[(\sin^2\theta_-^1-\sin^2\theta_+^1)\right . \nonumber\\
& - & \left . (\sin^2\theta_-^2-\sin^2\theta_+^2)\right]d\zbf k
\label{delrho}
\end{eqnarray}
$\rho^i=\langle {\psi^i}^\dagger\psi^i
\rangle$ , where the expectation value is taken with respect to the state given 
in Eq.(\ref{ubt}). 

 Using the gap equation Eq.(\ref{gapeq}), the thermodynamic potential
given by Eq.(\ref{thpot}) can be rewritten as
\bearr
&&\Omega(\Delta,\bar\mu,\delta_\mu,\beta) \nonumber\\
&&=
\frac{2}{(2\pi)^3\beta}\int d\zbf k\left[\left(\bar\xi_--\bar\omega_-+
\frac{\Delta^2}{2\bar\omega_-}
+\bar\xi_+-\bar\omega_++
\frac{\Delta^2}{2\bar\omega_+}
\right) \right . \nonumber\\
&& \left .-\sum_{i}\bigg\lbrace \ln\left(1+e^{(-\beta\omega_-^{(i)})}\right)
+\ln\left(1+e^{(-\beta\omega_+^{(i)})}\right)\bigg\rbrace\right]. 
\label{thpotd}
\eearr

To compare the stability of various phases we compare the thermodynamic 
potentials of these phases with respect to that of normal matter.
 This can be obtained from 
Eq.(\ref{thpotd}) in the limit of
gap $\Delta\rightarrow 0$. We consider the difference in the thermodynamic
potentials between condensed phase and the normal matter as given by
\begin{widetext}
\bearr
&&\tilde\Omega(\Delta,\bar\mu,\delta_\mu,\beta)=
\Omega(\Delta,\bar\mu,\delta_\mu,\beta)-
\Omega(\Delta=0,\bar\mu,\delta_\mu,\beta)\nonumber\\
&&=
\frac{2}{(2\pi)^3}\int\left(|\bar\xi_-|-\bar\omega_-+
\frac{\Delta^2}{2\bar\omega_-}\cos 2\theta_-^{1,2}
+|\bar\xi_+|-\bar\omega_++
\frac{\Delta^2}{2\bar\omega_+}\cos 2\theta_+^{1,2}
\right)d\zbf k \nonumber\\
&&-\frac{2}{(2\pi)^3\beta}\sum_{i=1,2}\int\left[\big \lbrace \ln\left(1+e^{(-\beta\omega_-^{(i)})}\right)
+\ln\left(1+e^{(-\beta\omega_+^{(i)})}\right)\big \rbrace -\big \lbrace 
\ln\left(1+e^{(-\beta\omega_{0-}^{(i)})}\right)
-\ln\left(1+e^{(-\beta\omega_{0+}^{(i)})}\right)\big \rbrace \right]d\zbf k
\label{thpotdr}
\eearr
\end{widetext}
In the above, $\omega_{0\mp}^{(1)}=|\bar\xi_\mp|\mp\delta_\mu$, 
$\omega_{0\mp}^{(2)}=
|\bar\xi_\mp|\pm\delta_\mu$, correspond to the normal matter dispersion relations 
for the two species. For stability of the condensed phase $\tilde\Omega$ 
has to be negative with $\Delta$ and $\bar\mu$
determined from the gap equation Eq.(\ref{gapr}) and the number density equation
Eq.(\ref{rhob}). Further, one has to
ensure that the solution corresponds to a minimum and not a maximum.
In what follows we shall restrict ourselves to the case of zero temperature
only. As noted earlier we shall  consider here without loss of generality,
$\delta_\mu > 0$. Also  we consider a system of equal masses for the 
fermions so that $\delta_\epsilon=0$.
This leads to the possibility of quasi particle energy for species `1',
 $\omega_-^{(1)}$
or the quasi anti-fermion energy for species `2', 
$\omega_{+}^{(2)}$ becoming negative.
In that case the distribution functions given by Eq.(\ref{them}),
become Heaviside ($\Theta$) - functions
i.e. $\sin^2\theta^a=\Theta(-\omega_a)$. 
Further, using the identity
$\lim_{a\rightarrow \infty}\ln (1+\exp(-a x))/a=-x\Theta(-x)$, in 
Eq. (\ref{thpotdr}), the zero temperature thermodynamic potential
becomes
\begin{widetext}
\bearr
&&\tilde\Omega_0(\Delta,\bar\mu,\delta_\mu)=
\frac{2}{(2\pi)^3}\int\left(|\bar\xi_-|-\bar\omega_-+
\frac{\Delta^2}{2\bar\omega_-}
+|\bar\xi_+|-\bar\omega_++
\frac{\Delta^2}{2\bar\omega_+}
\right)
d\zbf k\nonumber\\
&+&
\frac{2}{(2\pi)^3}\int \left[\left(\omega_-^{(1)}-\frac{\Delta^2}{2\bar\omega_-}
\right)\theta(-\omega_-^{(1)})
+\left(\omega_+^{(2)}-\frac{\Delta^2}{2\bar\omega_+}\right)
\theta(-\omega_+^{(2)})-\omega_{0-}^{(1)}\theta(-\omega_{0-}^{(1)})
-\omega_{0+}^{(2)}\theta(-\omega_{0+}^{(2)})\right]d\zbf k
\label{thpotddr}
\eearr
\end{widetext}

The equations (\ref{gapr}) and (\ref{rhob})
need to be solved self consistently to determine the gap as a function 
of the coupling
and its stability will be decided by calculating the thermodynamic potential.
We note here that BEC is usually discussed using the canonical ensemble where
the particle number density is fixed as an external parameter.
We shall also consider  fixed number density here to discuss
 BCS--BEC crossover/phase transition
in the relativistic fermionic system  \cite{andreas,zhuangb,abukib,abuki,deng}.
 However, we might also note here that, to discuss quark matter,
one usually employs grand canonical ensemble with a fixed quark chemical potential
to explore the QCD phase diagram in the chemical potential and temperature plane.
In the numerical calculation that follows we keep the average number 
density fixed and consider the solutions as a function of the coupling and 
the difference in chemical potentials. 
Sometimes we find multiple solutions 
for the gap and average chemical potential satisfying Eq.s (\ref{gapr}) and 
(\ref{rhob}) corresponding to multiple extrema of the thermodynamic 
potential. In such cases, the solution which has the least 
thermodynamic potential is chosen. We also verify for the 
positivity for the second derivative of the thermodynamic potential
for this solution.
This way we ensure that the
pair of solution for superfluid gap and the chemical potential 
correspond to the minimum
of the thermodynamic potential.
The detailed numerical calculations of the present investigation are discussed
in the next section.

\section{Numerical solution of the gap equation and phase structure}
 For numerical calculations it is convenient to introduce the
dimensionless quantities in terms of Fermi momentum $k_f$
or Fermi energy $\epsilon_f=\sqrt{k_f^2+m^2}$, 
defined through $|\zbf k|=k_f x$, $\eta=1/(k_fa)$,
$m=k_f \hat m$ , $\Delta=\epsilon_f z$, $\mu=\epsilon_f \hat\mu$. 
The gap equation at zero 
temperature can then be written as
\begin{eqnarray}
-\frac{\eta}{2}&&=\frac{1}{\hat m\pi}\int_0^{x_{max}} dx x^2\left(
\frac{1}{\hat\omega_-}+\frac{1}{\hat\omega_+}-\frac{2 \epsilon(x)}
{x^2} \right . \nonumber\\
&& - \left . \frac{1}{\hat\omega_-^{(1)}}\Theta(-\hat\omega_-^{(1)})-
\frac{1}{\hat\omega_-^{(2)}}\Theta(-\hat\omega_-^{(2)})
\right).
\label{gapd}
\end{eqnarray}
Similarly, the equation for the average number density given by Eq.(\ref{rhob})
can be rewritten in terms of these dimensionless quantities as

\begin{eqnarray}
1&=&1.5\int_0^{xmax} dx x^2\left[\frac{\hat\xi_+(x)}{\hat\omega_+(x)}
\left(1-\Theta(-\hat\omega_+^{(2)})\right) \right . \nonumber\\
&& \left . -\frac{\hat\xi_-(x)}{\hat\omega_-(x)}
\left(1-\Theta(-\hat\omega_-^{(1)})\right)\right]
\label{rhod}
\end{eqnarray}
for the average density. Here, $\hat\xi_\pm(x)=\hat
\epsilon(x)\pm \bar{\hat\mu}\sqrt{(1+\hat m^2)}$, $\hat\epsilon(x)=\sqrt{x^2+\hat m^2}$,
$\hat\omega_\pm=\sqrt{\hat\xi^2_\pm+z^2(1+\hat m^2)}$ and finally $x_{max}=\Lambda/k_f$
is the upper cut-off for momentum in units of Fermi momentum $\k_f$. 
Here $\bar{\hat \mu}$ is the 
average of the chemical potentials of the two species in units of Fermi energy.

To analyze the crossover, let us first consider the symmetric case -- 
namely when the chemical potential difference between the two species is zero. 
In this limit, $\Theta(-\omega_-^{(i)})$ become zero in both the equations
(\ref{gapd}) and (\ref{rhod}). 
We note that we have essentially three dimensional quantities in the problem:
the cutoff $\Lambda$, the mass of the fermion $m$ and the
scattering length $a$. 
We further note that the dimensional coupling
$G$ is bounded above with a critical value $G_c\Lambda^2>2\pi^2$, beyond which
the zero density vacuum itself is unstable to form fermion pairs leading to a
Majarona mass for the fermions.

We might note here that at zero density, the minimum excitation 
energy for the fermion is its mass $m$. For a normal matter, with a finite
chemical potential it is
$(m-\mu)$. In the BEC state, the decay mode of the bound state is that of the
bound state going to two fermions. The threshold for this energy is 
thus $2(m-\mu)$.
The bosonic bound state should therefore be stable if this threshold energy 
is positive
which in turn means that $m>\mu$. This we shall take as our working definition for BEC
phase to distinguish between the BCS and BEC phase as we increase the 
dimensionless coupling
parameter $\eta$ from weak coupling BCS (large negative $\eta$) to strong coupling
BEC phase (large positive $\eta$) through unitary limit ($\eta$=0).
 

\begin{figure}
\includegraphics[width=8cm,height=8cm]{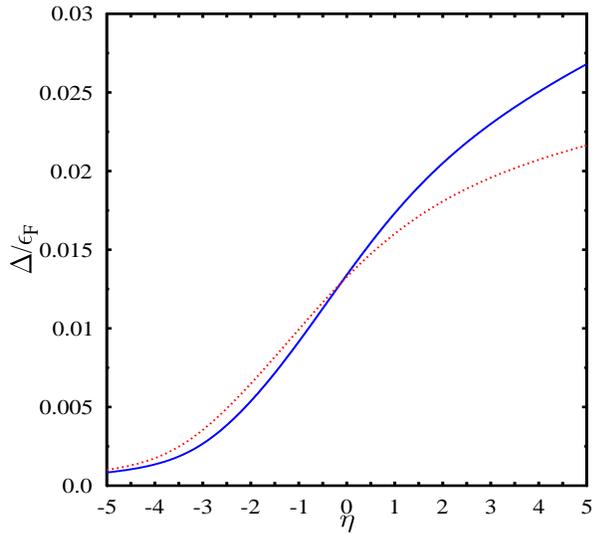}
\caption{\it  Gap parameter in units of Fermi energy 
is plotted as a function of the dimensionless coupling.
The dotted line corresponds to the case where anti particle contributions are
not included. The solid line corresponds to the case with inclusion of the 
antiparticle contributions.
In this plot, we have chosen $\Lambda/k_f=50$ and $m/k_f=5$}.
\label{fig1}
\end{figure}

\begin{figure}
\includegraphics[width=8cm,height=8cm]{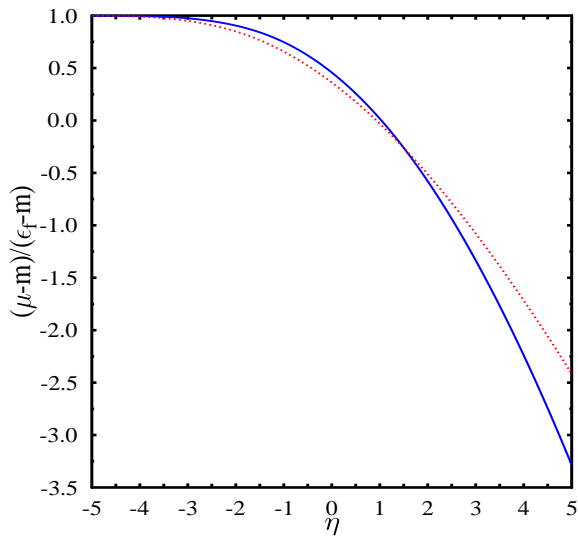}
\caption{\it 
The  scaled chemical potential $(\mu-m)/(\epsilon_f-m)$ is plotted as a function
of the dimensionless coupling, $\eta$. 
The dotted line corresponds to the case where anti particle contributions are
not included. The solid line corresponds to the case with inclusion of the antiparticle 
contributions.
In this plot, we have $\Lambda/k_f=50$ and $m/k_f=5$}.
\label{fig2}
\end{figure}

For numerical calculations, we  choose
the parameters $x_{max}=\Lambda/k_f =50$, $\hat m=m/k_f=5$
and study the results by varying the dimensionless coupling 
$\eta$ from negative to positive values. The resulting gap parameter and 
the chemical potential are plotted in figures {\ref{fig1}} and \ref{fig2}.
To show the contribution of the antiparticle 
degrees of freedom we have plotted the results 
obtained by solving the coupled gap equation Eq.(\ref{gapd}) 
and Eq.(\ref{rhod}), without and with the antiparticle contributions.

\begin{figure}[htbp]
\begin{center}
\includegraphics[width=8cm,height=8cm]{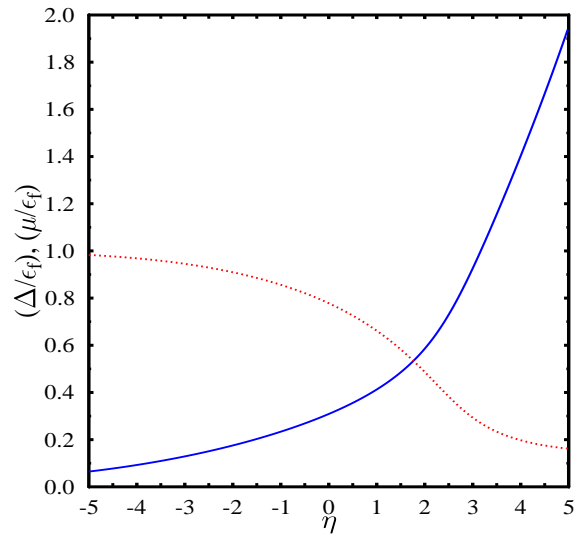}
\end{center}
\caption{\em{ Superfluid gap (solid line) and chemical potential (dotted line)
in units of Fermi energy as  functions of the dimensionless coupling, $\eta$.
}}
\label{fig3}
\end{figure}

One might naively expect that for this non relativistic $k_f/m=0.2$ case, the 
antiparticle channel
is suppressed. But as may be seen from Fig.{\ref{fig1}},
while such an expectation may be the situation for weak coupling BCS regime,
these contributions become increasingly important as the coupling increases. 
As the coupling increases, 
the chemical potential $\tilde\mu=\mu-m$ decreases and changes sign at coupling 
$\eta\approx 1.04$ signaling the BEC regime. To appreciate the relativistic 
effects we also consider the case with $m/k_f=0.67$ and $\Lambda/k_f=3.3$
\cite{andreas} and the resulting gap and the chemical potential are 
shown in Fig. \ref{fig3}.

As we might observe, in the weak coupling  BCS limit (large negative $\eta$),
the chemical potential is given by the Fermi energy. With increase in the coupling,
the chemical potential decreases and becomes negligible as compared to the Fermi energy
in deep BEC regime of large positive value of $\eta$.
The gap starts with exponentially small values in the weak coupling regime as 
expected from
BCS theory and rises monotonically as the coupling increases. It becomes of the 
order of the Fermi energy at around the unitary regime $\eta$=0.
 For $\eta=0$, the superfluid
gap and the chemical potential turn out to be $\Delta=0.3\epsilon_f$ and 
$\mu=0.78\epsilon_f$ respectively. We might note here that the same 
turns out to be about
$\mu=0.37\epsilon_f$ in Ref. \cite {andreas}. The discrepancy between the two 
can easily be 
understood by comparing the gap equation in the unitary limit in the two cases. The
difference lies in the way the renormalization of the gap equation is being 
done in 
these cases. To compare with the non relativistic results, we subtract out the
mass terms from both the chemical potential as well as the Fermi energy. The 
resulting ratio then becomes 
$\tilde\mu/\tilde\epsilon_f=(\mu-m)/(\epsilon-m)\approx 0.5$.
In the non relativistic fermionic models this value turns to be $0.4$ to $0.5$
\cite{amhmloff, nishidason,carlsonreddy,chennakano}. As $\eta$ increases, 
at about $\eta=1.68$, the chemical potential becomes smaller than 
the mass of the fermion
and the system goes to the BEC regime.

\begin{figure}
\begin{center}
\includegraphics[width=8cm,height=8cm]{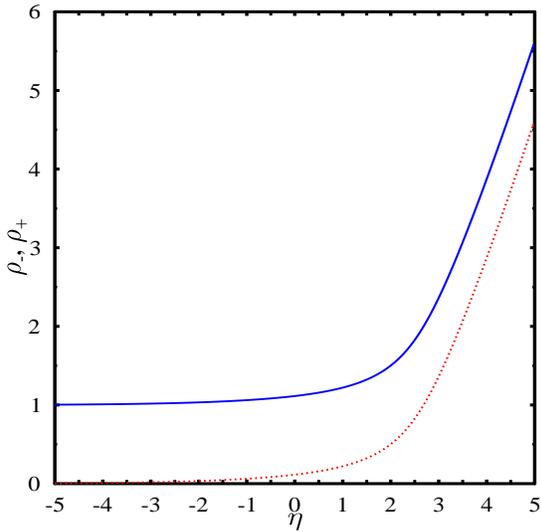}
\end{center}
\caption{\em{ Number densities of fermions, $\rho_-$ (solid line) and
 anti-fermions, $\rho_+$ (dotted line)
in units of $k_f^3/3\pi^2$ as a function of  dimensionless coupling $\eta$.
}}
\label{fig4}
\end{figure}

As the coupling becomes close to the unitary regime, the 
anti fermion contributions become important. In Fig. \ref{fig4}, 
we show the number densities of the
fermions $\rho_-$ and the anti-fermions $\rho_+$ as defined in Eq.(\ref{rhom})
and Eq.(\ref{rhop}) respectively.
While  in the BCS regime the antiparticle contribution
to the number density turns out to be negligible as compared to the particle
 contributions, it becomes increasingly important near the unitary regime.
 At very large values of $\eta$,
the chemical potential becomes negligible, the contributions of the 
particle and the antiparticle
to the number densities become large as compared to the density $\rho$ 
and their difference
produces a conserved net density \cite{zhuang}.

Next, we consider the case of superfluid with a mismatch in the 
chemical potentials
i.e. $\delta_\mu \neq 0$. In the numerical calculations, we keep the 
average density fixed and calculate the average chemical potential and the
superfluid gap using Eq.(\ref{rhob}) and Eq.(\ref{gapr}) respectively.
The stability of the solution is analysed by calculating the thermodynamic potential.
Sometimes,  particularly near the phase transition region,
there are multiple solutions of the gap and number density equations. 
Of these, we choose the solution which has the least 
thermodynamic potential. We also numerically verify the positivity of 
second derivative of the thermodynamic potential at this pair of 
$\Delta$ and $\bar\mu$. This way we
ensure that the solution we get is a minimum and not a maximum of the thermodynamic potential.

To analyze the nature of the solutions, we shall further assume, without 
loss of generality,
$\delta_\mu>0$. In that case, the quasi particle energy for species '1', 
$\omega_-^{(1)}
(\zbf k)= \bar\omega_-(\zbf k)-\delta_\mu$, and the same for 
quasi antiparticle energy of species '2', $\bar\omega_+^{(2)}(\zbf k)
=\bar\omega_+(\zbf k)-\delta_\mu$ can become negative. In that
case, at zero temperature the contributions of the $\Theta$ - functions 
both in the
gap equation Eq.(\ref{gapeq}), and, the thermodynamic potential Eq.(\ref{thpot})
can be non vanishing. The $\Theta$ - functions limit the range of the momentum 
integrations.  Solving for the zeros of $\omega_-^{(1)}$, we note that
this vanishes
at momenta $\zbf k_{min/max}^2=
\left ({\bar\mu} \pm \sqrt{\delta_\mu^2-\Delta^2} \right )^2
-m^2$. Similarly $\omega_+^{(2)}$ vanishes at momenta satisfying
$\zbf k_{min/max}^2=
\left (-{\bar\mu} \pm \sqrt{\delta_\mu^2-\Delta^2}\right )^2 - m^2$. 
Clearly this is possible provided the gap $\Delta$ is smaller
than $\delta_\mu$. The zeros of the dispersion relations correspond to 
effective Fermi surfaces. In general there can be two fermi surfaces
for species `1' along with the gapped ones. From the dispersion
relations for the quasi particles and antiparticles, it is clear that
we also can have the interesting possibility
of interior gap solutions for particles of species `1' and 
antiparticles of species `2'
\cite{andreas}.

\begin{figure}
\begin{center}
\includegraphics[width=8cm,height=8cm]{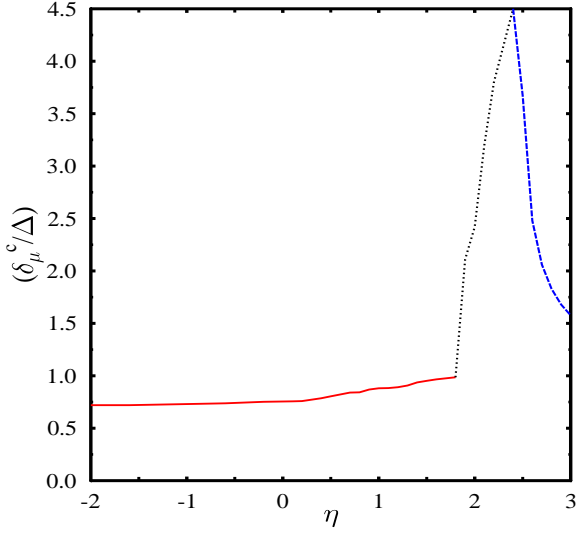}
\end{center}
\caption{\em{ Ratio of critical chemical potential difference to the gap
as a function of the coupling strength $\eta$. Gap less phase appear 
for $\eta>1.9$.
Solid line denotes the BCS regime and the dotted line indicates the regime 
where quasi particles of species `1' become gapless. The dashed line 
indicates the regime where the antiparticles of species `2' also become gapless.}}
\label{fig5}
\end{figure}

\begin{figure}
\begin{center}
\includegraphics[width=8.5cm,height=8.5cm]{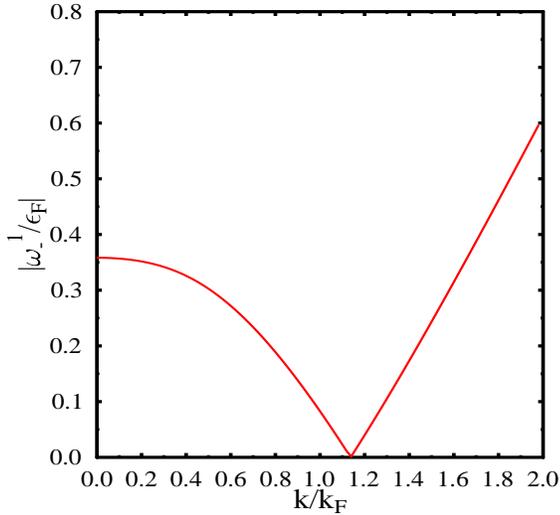}
\end{center}
\caption{\em{Quasi particle dispersion relation for species '1'. The plot is for
the case of coupling $\eta=2.1$ and chemical potential difference 
$\delta/\Delta_0=1.125$, $\Delta_0$ being the gap at zero chemical 
potential difference.}}
\label{fig6}
\end{figure}

After these general remarks regarding the topology of Fermi surfaces, let 
us discuss some
numerical results of the present investigation. As mentioned earlier, 
we keep the average density fixed and calculate the superfluid 
gap and the average chemical potential for a given coupling and 
a given chemical potential difference.
In case there are multiple nontrivial solutions for the gap and the 
average chemical potential, the solution for the gap and average chemical
potential pair is accepted which has the least thermodynamic potential.

In the numerical calculations that we present here, we take 
the parameters $\Lambda/k_f=3.3 $ and $m/k_f=0.67$.
In Fig. \ref{fig5}, we plot the
quantity $\delta_\mu^c/\Delta$, the ratio of maximum chemical potential 
difference to the superfluid gap, that can sustain pairing,
as a function of the dimensionless coupling $\eta$. For a larger value 
of $\delta_\mu$, beyond $\delta_\mu^c$ there are no acceptable solutions of
the gap and number density equation with a non vanishing gap and
with a lower thermodynamic potential. 
For weak coupling in the BCS limit,
it approaches the Clogston - Chandrasekhar limit of 
$\delta_\mu^c/\Delta\simeq 0.72$. 
Initially, as the coupling increases from BCS to BEC regime, $\delta_\mu^c/\Delta$
increases monotonically as shown by the solid line in figure \ref{fig5}.
In this regime there are no gapless modes and the density difference, 
$\delta_\rho$, 
between the two species is zero. At $\eta\simeq 1.9$, $\delta_\mu^c/\Delta$ 
increases sharply with $\eta$ until it reaches a value of the order of 2.2. 
This is shown by the dotted line in Fig.\ref{fig5}.
In this regime, $\omega_-^{(1)}$ becomes gapless while all other modes are
gapped. In this region, $(\bar\mu-m)$ is negative and hence this gapless
phase lies in the BEC regime. The density difference between the two species becomes nonzero. 
The dispersion relation $\omega_-^{(1)}(\zbf k)$  in this region is shown in Fig.\ref{fig6}.
In particular we have chosen here, $\eta=2.2$ and $\delta_\nu/\Delta_0=0.47$. 
Here $\Delta_0$ is the superfluid gap at zero chemical potential difference and turns
out to be $\Delta_0=0.637 \epsilon_f$. With these parameters, the 
average chemical potential turns out to be $\bar\mu=0.47\epsilon_f$ and 
the gap is $\Delta
\simeq 0.35\epsilon_f$.

\begin{figure}
\begin{center}
\includegraphics[width=8.5cm,height=8.5cm]{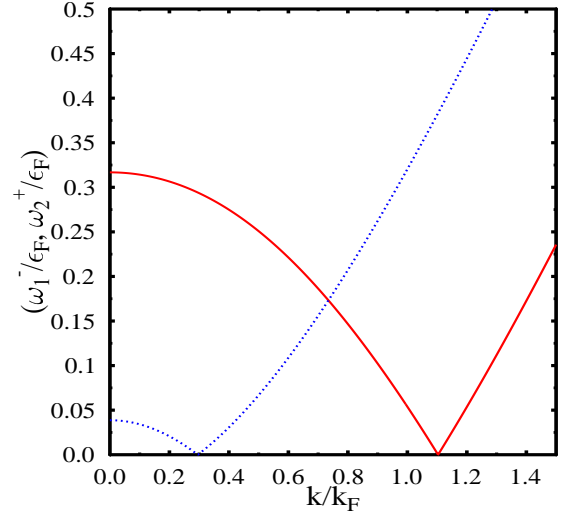}
\end{center}
\caption{\em{ Dispersion relation for gapless modes. Solid line shows 
the dispersion
relation for quasi particle of species '1'($\omega_-^{(1)}(\zbf k)$) 
and the dotted line
shows the dispersion relation for quasi-antiparticle of species '2'
($\omega_+^{(2)}(\zbf k)$). The plot is for the case of coupling $\eta=3$ and
$\delta_\nu/\Delta_0=1.195$.}}
\label{fig7}
\end{figure}

The ratio $\delta_c/\Delta$ increases with $\eta$ till $\eta\simeq 2.5$. 
Beyond this, 
there are two gapless modes. Both the usual quasi particle for species '1' and
the quasi anti particles of species `2' become gapless beyond this coupling.
The 
dashed line in Fig.\ref{fig5} represents this regime. In Fig. \ref{fig7} we show
the dispersion
relation for these gapless modes. We have taken the coupling $\eta=3$ and
$\delta/\Delta_0=1.195$ here. The values for the gap and the average chemical 
potential here is $\Delta=0.72\epsilon_f$ and ${\bar\mu}=0.23\epsilon_f$ respectively.

\begin{figure}
\begin{center}
\includegraphics[width=9.0cm,height=9.0cm]{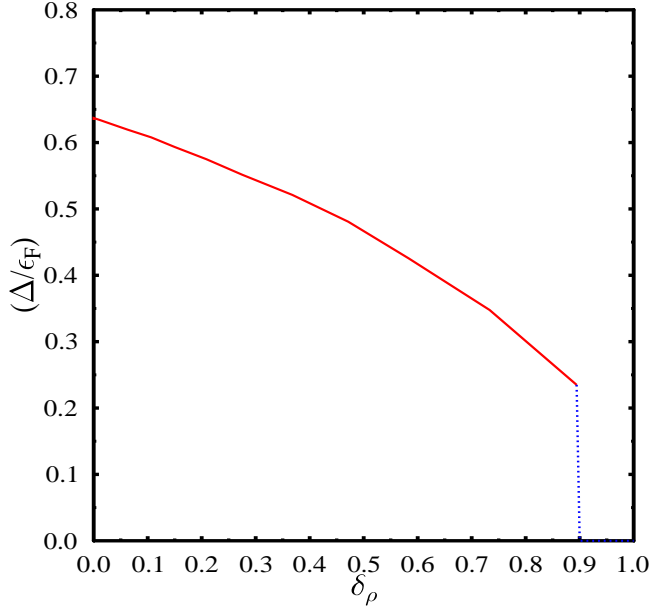}
\end{center}
\caption{\em{Superfluid gap as a function of difference in number densities
of the condensing species. This is plotted for $\eta=2.1$.}}
\label{fig8}
\end{figure}

\begin{figure}
\begin{center}
\includegraphics[width=8cm,height=8cm]{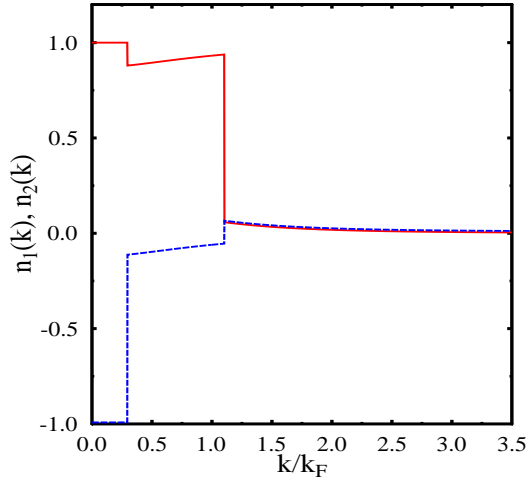}
\end{center}
\caption{\em{Number density distribution for the two species. Solid line
corresponds to species '1' and the dashed line corresponds to species '2'.
This plot corresponds to $\eta=3.0$ and $\delta_\mu/\Delta_0=1.195$}}.
\label{fig9}
\end{figure}
We also would like to note here that we did not find any breached pairing,
i.e.,  two Fermi surfaces with nonzero value of the gap for any value of 
the coupling $\eta$.
These results are similar qualitatively to those of Ref. \cite{andreas}
where a different renormalization was adopted for the crossover parameter.

We might mention here that we do not keep the density difference fixed
in our calculations. We perform the calculations for a fixed average density and a given chemical potential difference. 
In presence of gapless phases, the density difference becomes nonzero.
In Fig. \ref{fig8}, the dependence of the gap on the density difference, 
$\delta_\rho$ is shown for coupling $\eta=2.1$. Superconductivity is supported 
for a maximum density difference of $\delta_\rho\simeq 0.9 (k_f^3/3\pi^2)$
beyond which the system goes over to normal matter with zero gap.
For coupling $\eta < 1.9$ we do not find any superfluid
phase free energetically favorable with any non zero value of density
difference $\delta_\rho$. However, a chemical potential difference can
still support a Cooper paired state.

Next, let us consider the number density distribution
of the two species in the momentum space when gapless modes exist.
Let us note that the number densities at zero temperature for the two species 
are given by
\bearr
\rho_1 &=&\langle\psi_1^\dagger\psi_1\rangle\nonumber\\ &=&
\frac{2}{(2\pi)^3}\int d\zbf k\bigg(\sin^2 
f(\zbf k)+\Theta(-\omega_-^{(1)})\cos^2 f(\zbf k)\nonumber\\
 &-&\sin^2f_1(\zbf k)(1-\theta(-\omega_+^{(2)}))
\bigg)\nonumber\\ 
&\equiv&
 \frac{2}{(2\pi)^3}\int d\zbf k n_1(\zbf k)
\label{n1k}
\eearr
for species `1' and 
\bearr
\rho_2 &=&\langle\psi_2^\dagger\psi_2\rangle\nonumber\\& = &
\frac{2}{(2\pi)^3}\int d\zbf k\bigg(\sin^2 
f(\zbf k)(1-\Theta(-\omega_-^{(1)}))\nonumber\\& -&
(\sin^2f_1(\zbf k)+\cos^2 f_1(\zbf k)
\theta(-\omega_+^{(2)}))
\bigg)\nonumber\\
&\equiv&
 \frac{2}{(2\pi)^3}\int d\zbf k n_2(\zbf k)
\label{n2k}
\eearr
for species `2'.
Here, $\tan 2 f(\zbf k)=\Delta
 / \bar\xi_-$ and 
$\tan 2 f_1(\zbf k)=\Delta$ /
 $\bar\xi_+$.
In Fig. \ref{fig9} we have plotted the momentum space density distributions
$n_1(\zbf k)$ and $n_2(\zbf k)$  for $\eta=3.0$ and $\delta_\nu/\Delta_0=1.195$.
In this case both $\omega_-^{(1)}$ and $\omega_+^{(2)}$ become gapless. 
In the region where 
both $\omega_-^{(1)}$ and $\omega_+^{(2)}$ are negative, 
$n_1(k)=1$ and $n_2(k)=-1$ as may be seen from
equations (\ref{n1k}) and (\ref{n2k}). In the region,
 where only $\omega_-^{(1)}$ is negative,
$n_1(k)=(1+\bar\xi_+/\bar\omega_+)/2$ and 
$n_2(k)=-(1+\bar\xi_+/\bar\omega_+)/2$.
Finally, when both $\omega_-^{(1)}$ and $\omega_+^{(2)}$ are positive, 
both $n_1(k)$ and 
$n_2(\zbf k)$ are identical and are given by the relativistic BCS distribution function 
$n(\zbf k)=(\bar\xi_+/\bar\omega_+ -\bar\xi_-/\bar\omega_-)$. This is precisely
what is depicted in Fig.\ref{fig9}. Let us note here that although  the
individual distribution functions in the momentum space for the two species
could be negative, the average occupation number densities $\bar n(\zbf k)=
(n_1(\zbf k)+n_2(\zbf k))/2$ as well as the difference in occupation number densities
$\delta_n(\zbf k)=(n_1(\zbf k)-n_2(\zbf k))/2$ are always positive definite.

Most of the results obtained in this section for the model with a 
four fermion interaction are similar to the mean field results obtained
in a boson-fermion model \cite{andreas}.
It is nice to see the similarity to the mean field results of Ref.
\cite{andreas} which in our investigation arises with a simple ansatz
for the ground state given by
of Eq. (\ref{ubt}) determined through an extremisation of the 
thermodynamic potential.
As emphasized in section \ref{model}, the scalar condensate field
was considered as a classical auxiliary field. In the following 
section we shall treat them as dynamical fields and generalize the 
ansatz of Eq.(\ref{ubt}) to include the quanta of this field along
with those of the fermions. We shall illustrate this for the 
symmetric case i.e. when there is no chemical potential mismatch
for the two species of fermions.

\section{dynamical condensate fields and generalization of the BCS ansatz}

As discussed earlier, we shall now treat the the scalar filed $\Phi$ introduced in Eq.(\ref{liphi})
as  a dynamical field and rewrite the Lagrangian as
\be
{\cal L}={\cal L}_f+{\cal L}_b+{\cal L}_{bf}
\ee
where,
\be
{\cal L}_f= \bar\psi^i(i\gamma^\mu\partial_\mu-m+\mu\gamma^0)\psi^i,
\ee
\bearr
{\cal L}_b&=& (\partial_0-i\mu_B)\Phi^\dagger(\partial_0+i\mu_B)\Phi
-m_b^2\Phi^\dagger\Phi \nonumber\\&-&(\nabla\Phi^\dagger)(\nabla\Phi)
-\lambda(\Phi^\dagger\Phi)^2
\label{lb}
\eearr
and,
\be
{\cal L}_{bf}= 
g|\epsilon^{ij}|\left(\bar\psi^i\gamma^5\psi^j_c\Phi+
\bar\psi^i_c\gamma^5\psi^j\Phi^* \right).
\label{libf}
\ee
We shall illustrate the effect of the dynamical bosonic field on the 
BCS-BEC crossover physics and hence will consider the case where there is
no mismatch in the chemical potentials of the two condensing fermioionic 
species with a common chemical potential $\mu$..
For the dynamical bosonic field now we have introduced the chemical potential
$\mu_B$ which is equal to twice the fermionic chemical potential 
$\mu$ in equilibrium.
We have also included a quartic term in the scalar field in ${\cal L}_B$ for the sake of completeness.
In fact, this quartic term leads to nonperturbative corrections to
the thermodynamic potential 
as we shall observe later. Clearly, in the absence of the kinetic terms and the
quartic interaction term, this Lagrangian reduces to the one considered in section \ref{model}.

As before, we shall consider a state such that $\langle \Phi\rangle=\phi_0=\langle \Phi^\dagger\rangle$
and investigate the fluctuations of the condensate field by
defining the quantum fields
$\Phi'=\Phi-\phi_0$, and $\Phi'^\dagger
=\Phi^\dagger-\phi_0$.  Then ${\cal L}_b$ reduces to
\bearr
{\cal L}_b&\simeq& (\partial_0-i\mu_B)\Phi'^\dagger(\partial_0+i\mu_B)\Phi'
-m_b^2\Phi'^\dagger\Phi'\nonumber\\&-&(\nabla\Phi'^\dagger)(\nabla\Phi')
-\lambda (\Phi'^\dagger\Phi')^2 -4\lambda\phi_0^2(\Phi'^\dagger\Phi')\nonumber\\
&-&V_0(\phi_0),
\label{lbb}
\eearr
where, $\simeq$ means neglecting odd powers in $\Phi'$s. Further, in the above,
$V_0(\phi_0)$ is the 'tree' level potential given as
\be
V_0(\phi_0)=(m_b^2 -\mu_B^2)\phi_0^2+\lambda\phi_0^4
\label{v0}
\ee
One can write down the corresponding Hamiltonian densities as
\be
{\cal H}={\cal H}_f+{\cal H}_b+{\cal H}_{bf}
\label{ham}
\ee
with
\be
{\cal H}_f=\sum_{i}\psi_i^\dagger(-i\zbf\alpha\cdot\zbf\nabla +\beta m)\psi_i
\ee
\bearr
{\cal H}_b&=&
\Pi_{\Phi'^\dagger}\Pi_{\Phi'}+i\mu_b(\Pi_{\Phi'}{\Phi'}\nonumber\\
&-&\Pi_{\Phi'^\dagger}{\Phi'^\dagger})
+\Phi'^\dagger(-\zbf\nabla^2+m_b^2)\Phi' \nonumber\\
&+& \lambda (\Phi'^\dagger\Phi')^2 +4\lambda\phi_0^2(\Phi'^\dagger\Phi')+V_0(\phi_0)
\label{hb}
\eearr
and ${\cal H}_{bf}=-{\cal L}_{bf}$. Here, $\Pi_{\Phi'}$ ($\Pi_{\Phi'^\dagger})$ is the conjugate momentum of the
corresponding field $\Phi'$ ($\Phi'^\dagger$). Similar to the fermion field operator expansion we can take the boson
field operator expansions in terms of creation and annihilation operators. We have thus e.g. 
\bearr
&&\Phi'(\zbf x,t=0)\nonumber\\&&=\frac{1}{(2\pi)^{3/2}}\int d \zbf k \frac{1}{\sqrt{2\omega(k)}}(a(\zbf k)+b^\dagger(-\zbf k))e^{i \zbf k\cdot\zbf x}
\eearr
and
\bearr
&&\Pi_{\Phi'}(\zbf x,t=0)\nonumber\\&&=\frac{i}{(2\pi)^{3/2}}\int d \zbf k 
\sqrt{\frac{\omega(k)}{2}}(-b(\zbf k)+a^\dagger(-\zbf k))e^{i \zbf k\cdot\zbf x}
\eearr
Let us note that with the above expansion for the conjugate fields satisfying the quantum algebra $[\Phi'(\zbf x),\Pi_{\Phi'}(\zbf y)]=
i\delta(\zbf x-\zbf y)$ leads to the usual commutation relations for the creation and annihilation operators
$[a(\zbf k),a^\dagger(\zbf k')]= \delta(\zbf k-\zbf k')=
[b(\zbf k),b^\dagger(\zbf k')]$ for {\em any arbitrary function $\omega(\zbf k)$}.

With the operators for the scalar fields defined, we now geneneralise the ansatz given in Eq.(\ref{ubt}) and write down the 
ansatz $|\Omega(\beta,\mu\rangle_B$  to include the effects of the boson field
as,
\begin{equation} 
|\Omega(\beta,\mu)\rangle_B={\cal U}_{\beta,\mu}^B
U^B|\Omega(\beta,\mu)\rangle,
\label{ubbt}
\end{equation} 
Here, similar to 
Eq.(\ref{ubt}) and Eq.(\ref{bd}) the operator $U^B$ is given as,
\be
U^B=\exp\left(\int d\zbf k g(k) a^\dagger(\zbf k)b^\dagger(-\zbf k)-h.c.\right)
\label{ubb}
\ee
and similar to Eq.(\ref{ubm}) and Eq.(\ref{bth}), the operator corresponding
to thermal excitations of the bosonic
fields ${\cal U}^B_\beta\mu$ is given as
 \bearr
{\cal U}_{\beta\mu}^B
&=&\exp\bigg(\int d\zbf k \bigg(a'^\dagger(\zbf k)\underline a(-\zbf k) \theta_a(\zbf k,\mu)\nonumber\\
& + &
b'^\dagger(\zbf k)\underline b(-\zbf k) \theta_b(\zbf k,\mu)\bigg)
-h.c\bigg).
\label{bbth}
\eearr

This leads to e.g.
\bearr
&&\langle\Phi'(\zbf x)\Phi'(\zbf y)\rangle =\frac{1}{(2\pi)^3}\int\frac{d\zbf k}{2\omega(\zbf k)}e^{i\zbf k\cdot(\zbf x-\zbf y)}\nonumber\\
&&
\bigg[(\cosh 2 g(\zbf k)+\sinh 2 g(\zbf k))
(\cos h^2\theta^a+\sin h^2\theta^b)\bigg]\nonumber\\ 
&\equiv& I(\zbf x-\zbf y, \beta)
\label{ixt}
\eearr
where, the expectation value is taken with respect to the state
$|\Omega(\beta,\mu)\rangle_B$ defined in Eq.(\ref{ubbt}).
The extra functions $g(\zbf k)$ as well as the thermal bosonic functions $\theta^{a,b}$ are to be determined as
earlier by extremisation of the thermodynamic potential. The thermodynamic potential for the boson fermion system
can be written as
\be
\Omega_{tot}=\Omega + \Omega_B
\label{omtot} 
\ee
Here, the fermionic contribution $\Omega$  has already been evaluated in Eq.(\ref{omgx}). The bosonic contribution $\Omega_B$
is given by
\be
\Omega_B=\epsilon_B-\mu_B\rho_B-\frac{1}{\beta}s_B.
\label{omgb}
\ee
The contribution of the first two terms is just the expectation value of Eq.(\ref{hb}) with respect to the state in Eq.(\ref{ubb}). The bosonic
entropy density $s_B$ is given similar to their
fermionic counterpart in Eq.(\ref{ent}), as
\cite{tfd},
\bearr
s_B &=& \frac{1}{(2\pi)^3}\sum_{i}\int d \zbf k
\Big (\cosh^2\theta_a\ln \cosh^2\theta_a \nonumber\\ &-&
\sinh^2\theta_a\ln \sinh^2\theta_a+ a\rightarrow b \Big).
\label{entb}
\eearr
Extremisation of the total thermodynamical potential with respect to 
the fermionic functions $f(\zbf k)$, $f_1(\zbf k)$ and $\theta_\mp^i(\zbf k)$
leads to the same
solutions for them as given in section \ref{evaluation}.  Extremising
the bosonic function $g(\zbf k)$ leads to
the solution
\be
\tanh 2g(k)=\frac{\omega^2-\zbf k^2-M^2} {\omega^2-\zbf k^2+M^2} 
\ee

with the quantity $M^2$ satisfying the temperature dependent 
'mass gap' equation given as 
\bearr
&&M^2=m^2+4\lambda\phi_0^2
\nonumber\\
&+&\frac{4\lambda}{(2\pi)^3}\int\frac{d\zbf k}{2\sqrt{\zbf k^2+M^2}}
(\cosh^2\theta_a+\sinh^2\theta_b)\nonumber\\
&=&m^2+4\lambda(\phi_0^2+I(\beta)).
\label{mgap}
\eearr
Here, $I(\beta)=I(\zbf 0,\beta)$, as given in Eq.(\ref{ixt}). Similarly,
minimizing the thermodynamic potential with respect to the bosonic thermal functions
yields
\be
\sinh^2\theta_a=\frac{1}{\exp(E_B-\mu_B)-1}\equiv n_B(\zbf k)
\ee
for the boson particle distribution function and
\be
\sinh^2\theta_b=\frac{1}{\exp(E_B+\mu_B)-1}\equiv n_{\bar B}(\zbf k)
\ee
for the boson anti-particle distribution function and
$E_B=\sqrt{\zbf k^2+M^2}$, with the 
temperature dependent mass $M$ satisfying the self consistent mass gap equation Eq.(\ref{mgap}).

With all the functions in the ansatz state Eq.(\ref{ubbt}) now determined, the total thermodynamic potential
can be written as, using Eq. (\ref{thpot}) and Eq.(\ref{omtot}),
\be
\Omega_{tot}=\Omega_f+\Omega_B
\ee
where, the fermionic part of the thermodynamic potential is given by
\bearr
\Omega_f&=&
\frac{2}{(2\pi)^3}\int\left(2\epsilon(k)-\bar\omega_--\bar\omega_+\right)
d\zbf k\nonumber\\
&-&
\frac{2}{(2\pi)^3\beta}\sum_i\int d\zbf k \bigg[\ln\left(1+e^{(-\beta\omega_-^{(i)})}\right)
\nonumber\\
&+&\ln\left(1+e^{(-\beta\omega_+^{(i)})}\right)\bigg].
\label{thpotf}
\eearr
As compared to expression in Eq.(\ref{thpot}), the above differs by the mass term $m_b^2\phi_0^2$ 
which is absorbed naturally in the bosonic part of the thermodynamic potential and the later
is given as
\bearr
\Omega_B &=&
 \frac{1}{(2\pi)^3}\int d\zbf k
\sqrt{\zbf k^2+M^2}\nonumber\\
 &-& 2 \lambda I^2(\beta)+
\frac{1}{\beta(2\pi)^3}\int d\zbf k
\sum_{i=1}^2\ln(1-\exp(E_i))\nonumber\\
&+&V_{0},
\label{omgbb}
\eearr
where,the summation is over the bosons and antibosons with $E_1=E_B-\mu_B$,
$E_2=E_B+\mu_B$,
and, $V_0$ is the tree level potential 
\be
V_0=(m_b^2-\mu_B^2)\phi_0^2+\lambda\phi_0^4
\label{vv0}
\ee
Finally, the extremisation of the total thermodynamic potential with respect to $\phi_0$ leads the superconducting gap equation
\be
M^2-\mu_B^2
=2\lambda\phi_0^2+4 g^2
\int\frac{d\zbf k}{(2\pi)^3}\left[
\frac{\cos 2\theta_-^{1,2}}{\bar\omega_-}+\frac{\cos 2\theta_+^{1,2}}{\bar\omega_+}
\right],
\label{gapeq1}
\ee
with, the mass $M$ satisfying the mass gap equation Eq.(\ref{mgap}) 
and 
$\cos 2\theta_{\mp}^{1,2}=(1-\sin^2\theta_{\mp}^1(\zbf k)-\sin^2\theta_{\mp}^2(\zbf k))$
with  $\sin^2\theta_\pm^i(\zbf k)$ being the thermal distribution functions
for the fermions defined in Eq.(\ref{them}).
This is the parallel of Eq.(\ref{gapeq}) where the condensate field was considered as an auxiliary field and 
there was no quartic coupling term for the scalar field.

The expression for the bosonic part of the thermodynamic potential
$\Omega_B$ in Eq.(\ref{omgbb}), is, however, affected by
two types of divergences, one arising from
 the divergent integrals as vacuum terms ($\phi_0=0$ at $T=0,\mu=0$) 
 which are still to be subtracted, and, the other being the
fact that the mass parameter $M^2$ is not well defined because 
of the infinities in the mass gap equation Eq.(\ref{mgap}). 
This can be taken care of by defining the renormalized quartic 
coupling and the boson mass through the relations \cite{cameliapi,rislena}
\begin{mathletters}
\be
\frac{1}{\lambda_R}= \frac{1}{\lambda}+4 I_2(\Lambda,\mu_{sc})
\label{lamr}
\ee
\be
\frac{m_R^2}{\lambda_R}= \frac{m_b^2}{\lambda}+4 I_1(\Lambda,\mu_{sc})
\label{mr}
\ee
\end{mathletters}
where, $I_1$ and $I_2$ are divergent integrals evaluated with a three momentum cutoff
defined as

\begin{mathletters}
\be
I_1=
 \frac{1}{(2\pi)^3}\int^{|\zbf k|<\Lambda} \frac {d \zbf k}{2|\zbf k|}
=\lim_
 {\Lambda\rightarrow
\infty}
\frac{\Lambda^2}{8\pi^2}
\label{i1}
\ee
and,
\bearr
&&I_2=
 \frac{1}{\mu_{sc}^2(2\pi)^3}\int^{|\zbf k|<\Lambda}  {d \zbf k}
\left(\frac{1}{2|\zbf k|}-\frac{1}{2\sqrt{\mu_{sc}^2+\zbf k^2}}\right)\nonumber\\
&=&\frac{1}{16\pi^2}\left(\ln\left(\frac{4\Lambda^2}{\mu_{sc}^2}\right)-1\right),
\label{i2}
\eearr
\end{mathletters}
where, $\mu_{sc}$ is the renormalization scale and $\Lambda$ is the three momentum cutoff.
Using the renormalization parameters the mass gap equation can be written in terms of finite
quantities as
\be
M^2=m_R^2+4\lambda_R(\phi_0^2+I_f(\beta))
\label{mgapr}
\ee
with $I_f(\beta)$ given as
\bearr
&&I_f(\beta) =\frac{M^2}{16\pi^2}
\left(\ln\left( \frac{M^2}{\mu_{sc}^2}\right)+1\right)\nonumber\\
&+&
\int\frac{d\zbf k}{(2\pi)^3} \frac{1}{2E}\left(n_B(\zbf k)+n_{\bar B}(\zbf k)\right).
\label{pfb}
\eearr

Similarly the bosonic part of the effective potential Eq.(\ref{omgbb}) in terms of the renormalized parameters,
subtracting out the vacuum terms becomes finite and is given as
\be
\Omega_B=V_0+V_1+V_2
\ee
with 
\be
V_0=m_R^2\phi_0^2+\lambda_R\phi_0^4-\mu_B^2\phi_0^2 +(\lambda_R-\lambda)\phi_0^4
\label{vvv0}
\ee
\be
V_1=
\frac{1}{\beta(2\pi)^3}\int d\zbf k
\sum_{i=1}^2\ln(1-\exp(E_i))
+\frac{M^4}{32\pi^2}\ln\left(\frac{M^2}{\mu_{sc}^2}\right)
\label{v1}
\ee
and,
\be
V_2=-2\lambda_R I_f(\beta)^2
\label{v2}
\ee

\noindent The cut off dependence in the effective potential is still there in the 
last term of Eq. (\ref{vvv0}) which disappears
in the limit of $\Lambda\rightarrow\infty$ as the bare coupling $\lambda$ vanishes in 
that limit. In the present calculations, however,
we keep the cutoff finite. We might note here that the difference between the bare and 
the renormalized quartic coupling can be written
as
\be
\lambda-\lambda_R=\frac{4\lambda_R I_2(\Lambda,\mu_{sc})}{1-4\lambda_R I_2(\Lambda,\mu_{sc})}\lambda_R
\ee
\noindent 
where, $I_2$ is given in Eq.(\ref{i2}). Mostly in our numerical calculations we shall have the limit $4\lambda_RI_2<<1$ in which case,
the contribution of the last term in Eq.(\ref{vvv0}) is negligible.

Next, the fermionic part of the thermodynamic potential Eq.(\ref{thpotf}) and the terms $V_0$, $V_2$ of the bosonic part can be combined using the
gap equation Eq.(\ref{gapeq1}) to yield a form similar to Eq. (\ref{thpotd}) as,
\bearr
&&\Omega^1_f\equiv\Omega_f+V_0+V_2\nonumber\\
&=&
\frac{2}{(2\pi)^3\beta}\int d\zbf k\left[\left(\bar\xi_--\bar\omega_-+
\frac{\Delta^2}{2\bar\omega_-}
+\bar\xi_+-\bar\omega_++
\frac{\Delta^2}{2\bar\omega_+}
\right) \right . \nonumber\\
&& \left .-\sum_{i}\big \lbrace \ln\left(1+e^{(-\beta\omega_-^{(i)})}\right)
+\ln\left(1+e^{(-\beta\omega_+^{(i)})}\right)\big \rbrace\right]\nonumber\\
&-&\lambda_R\phi_0^4 -(\lambda_R-\lambda)\phi_0^4 -2\lambda_R I_f(\beta)^2. 
\label{thpotdf}
\eearr

The number density equation, Eq.(\ref{rhob}) 
now gets modified due to the presence of dynamical condensate fields 
having a chemical potential twice as that of the fermions and the number density equation becomes

\bearr
\bar\rho&=&
8\mu\phi_0^2+2\int \frac{d\zbf k}{(2\pi)^3}
\left(\sinh^2\theta_a-\sinh^2\theta_b\right)\nonumber\\
&+& \frac{1}{(2\pi)^3}\int
\left[\frac{\xi_+}{\bar\omega_+}\cos 2\theta_+^{1,2}
-\frac{\xi_-}{\bar\omega_-}\cos 2\theta_-^{1,2}\right]d\zbf k
\label{rhot}
\eearr
To discuss the crossover, we define the mass parameter $m_1$ as in Ref.
\cite{andreas} as
\be
m_1^2=m_R^2-4g^2\int \frac{d\zbf k}{(2\pi)^3} \frac{2}{\sqrt{\zbf k^2+m^2}}
\label{m1}
\ee
and define the crossover parameter as
\be
x=-\frac{m_1^2-\mu_B^2}{4g^2}
\label{x}
\ee
The gap equation Eq.(\ref{gapeq1}) becomes
\bearr
m_1^2-\mu_B^2&=&2(\lambda-\lambda_R)\phi_0^2-2\lambda_R\phi_0^2-4\lambda_R I_f(\beta)\nonumber\\
&+&4g^2
\int\frac{d\zbf k}{(2\pi)^3}\bigg [
\frac{\cos 2\theta_-^{1,2}}{\bar\omega_-}+
\frac{\cos 2\theta_+^{1,2}}{\bar\omega_+}\nonumber\\
&-&\frac{2}{\sqrt{\zbf k^2+m^2}}
\bigg].
\label{gapeq2}
\eearr

Let us note here that the superfluid  gap equation has now contributions 
from the bosonic fluctuations through the
contribution $I_f(\beta)$. For numerical calculations, we choose a given value of the bosonic mass parameter
$m_R$  and solve the coupled number density equation and the superfluid gap equation
Eq.(\ref{gapeq2}) for the chemical potential and the superfluid gap. At each stage of evaluation of the
rhs of Eq.(\ref{rhot}) and Eq.(\ref{gapeq2}), the boson mass parameter 
$M^2$ is calculated self-consistently solving
Eq.(\ref{mgapr}). Throughout the numerical calculations we have chosen $g=2 \sqrt 2$ and
cutoff scale $\mu_{sc}^2=m_R^2$. 
This way we obtain the superfluid gap and the chemical potential for a given value of $m_R$  or equivalently,
for a given value of $x$ obtained through Eq.(\ref{m1}) and  Eq.(\ref{x}). These are then used to calculate any 
other thermodynamical quantities. 
\begin{figure}
\begin{center}
\includegraphics[width=12.0cm,height=6.86cm]{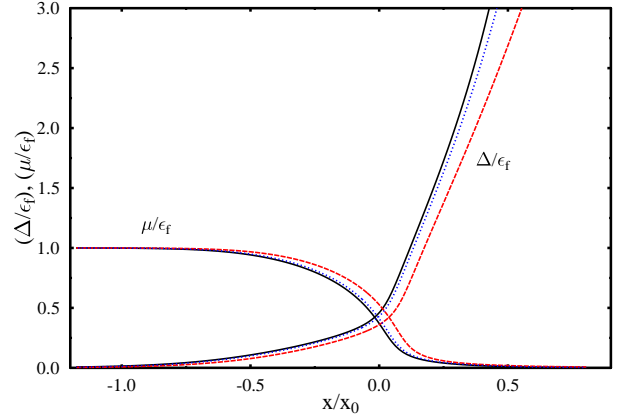}
\end{center}
\caption{\em{Fermion chemical potential and gap in units of fermi energy as a function of
dimensionless crossover parameter ${x}/{x_0}$ . 
 The solid (black), dotted (blue) and dashed (red) curves correspond to $\lambda_R$=0, 0.5 and 2 respectively.}}
\label{fig10}
\end{figure}
 In Fig.(\ref{fig10}), we show the numerical solutions of such a calculation for the gap 
 and the chemical potential
for zero temperature as a function of the dimensionless order parameter
$x/x_0$ with $x_0=\Lambda^2/(4\pi^2)$
for three values of the quartic coupling $\lambda_R$=0, 0.5 and 2.0. $\lambda=0$, will correspond to the mean field results.
As we might observe, the effect of the bosonic fluctuations are almost negligible in the BCS regime ($x/x_0<0)$ as well as near the unitary
regime $(x/x_0\sim 0)$. The effects of the fluctuating field is seen at 
large values of the crossover parameter in the deep BEC regime, manifesting 
in a
small reduction of the superfluid gap. The magnitude of this reduction increases with the quartic coupling.
This is due to the following reason. It is clear from observing the gap equation Eq.(\ref{gapeq2})
that, 
the effect of the positive quartic coupling leads to an increase
in the bosonic chemical potential for the same value of the gap parameter.
This leads to an increase of the corresponding value of the crossover
parameter '$x$' as may be clear from eq.(\ref{x})  for the same superfluid gap parameter.
This is also reflected in the number densities of
the fermions and bosons which we show in Fig.(\ref{fig11}). 
The change in chemical potential and the gap is appreciable from their
corresponding mean field values only in the BEC regime.
\begin{figure}
\begin{center}
\includegraphics[width=12.0cm,height=6.86cm]{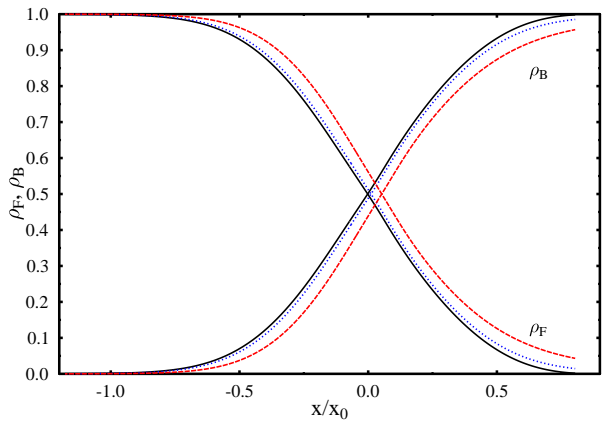}
\end{center}
\caption{\em{Number densities of fermions and bosons in units of total number density. The solid (black),
dotted (blue) and the dashed (red)
curves correspond to $\lambda_R$ =0,0.5 and 2 respectively. 
}}
\label{fig11}
\end{figure}

\begin{figure}
\begin{center}
\includegraphics[width=12.0cm,height=8cm]{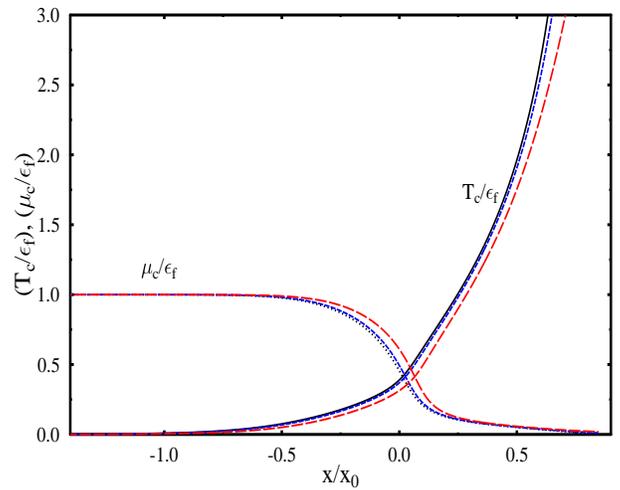}
\end{center}
\caption{\em{Critical temperature $T_c$ and the chemical potential
at $T=T_c$
in units of fermi energy as a function of
dimensionless crossover parameter . 
 The solid (black), dotted (blue) and dashed (red) curves correspond to $\lambda_R$=0, 0.5 and 2 respectively.}}
\label{fig12}
\end{figure}

 We then present the results for the critical temperature and the chemical potential as a function of the crossover parameter. This is done by setting
$\Delta=0=\phi_0$ in the number density equation Eq.(\ref{rhot}), 
in the gap equation Eq.(\ref{gapeq2}) as well as in the mass gap equation
Eq.(\ref{mgapr}) .
The results are shown in Fig. \ref{fig12}. While the behaviour of the chemical potential 
is qualitatively similar to that at zero temperature,
the critical temperature behaves similar to the gap at zero temperature 
with the crossover parameter.
The correction to the critical temperature becomes significant in 
the BEC regime only and increases with the quartic coupling. In this case,
again the reduction of the critical temperature lies due to the increase in 
chemical potential due to thermal as well as vacuum fluctuations as may be seen
in Eq.(\ref{gapeq2}). With further increase of the quartic coupling, 
the critical temperature become less and less steep due to larger
contributions of the thermal fluctuation of the the bosonic field.

\begin{figure}
\begin{center}
\includegraphics[width=12.0cm,height=8cm]{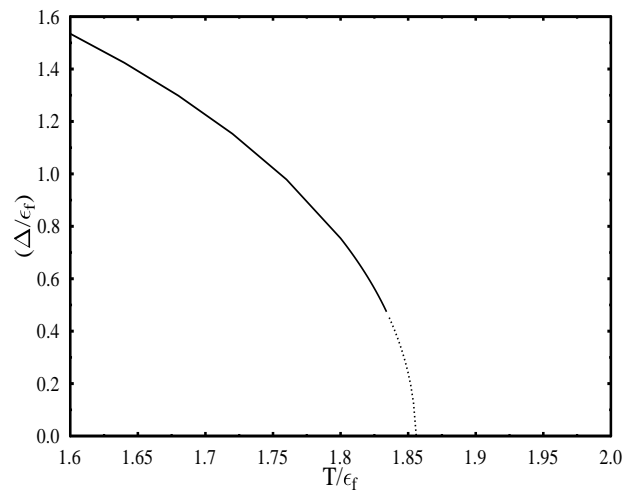}
\end{center}
\caption{\em{ Superfluid gap as a function of temperature for quartic 
coupling $\lambda_R$=5. The dotted line corresponds unstable solutions 
which are to solutions of the gap equation but with a higher thermodynamic
potential as compared to $\Delta =0$.}}
\label{fig13}
\end{figure}

To see the effect of the fluctuations of the condensate
field we also looked into the behavior of the gap as a function of
the temperature for different values of the quartic coupling
parameter $\lambda_R$. As the magnitude of the quartic coupling is increased, 
it was observed that the order parameter $\Delta$
changes discontinuously at the critical temperature. A typical behavior 
for the solution of gap equation Eq.(\ref{gapeq2}) is shown in Fig.\ref{fig13}
for $\lambda_R=5$. 
We note that near the critical temperature 
there are solutions
to the gap equation but having  larger values of the 
thermodynamic potential as compared to the normal matter. This is
suggestive of a first order phase transition when the effect of
fluctuations becomes large.
One should however be careful in drawing conclusions
from extrapolation to such a large value of the quartic coupling, 
because, although the result here is nonperturbative, it is
limited by the ansatz for the ground state in Eq.(\ref{ubbt}).
In this context we might remark here that gauge field fluctuations
in color superconductors changes the superconducting phase transition 
to a first order transition \cite{risph}. Similar observations were also made 
in Ref.\cite{deng} in a fermion-boson model where the fluctuations were 
treated within  a CJT formulation.

\section{summary}\label{summary}

We have considered here a variational approach to discuss the ground state structure of 
system of two species of relativistic fermions with a mismatch in their 
Fermi momenta. An explicit construct for the ground state is considered to
describe the two fermion condensates.
The ansatz functions including the distribution functions are determined by an 
extremisation of the thermodynamic potential.

 The quadratically divergent gap equation is made logarithmically divergent by
subtracting out the vacuum contribution and the four Fermi coupling is related
to the s-wave scattering length as in Ref.\cite{abuki,zhuang}. Unlike the
nonrelativistic case, the antiparticle degrees of freedom become important
even for the case $k_f/m<<1$, particularly for large values of 
$\eta (\equiv 1/k_fa)$.

When the Fermi momenta of the two species are different, we do not observe 
any gapless modes in the BCS regime. A nonzero chemical potential difference
can support a uniform BCS pairing with zero number density difference 
between the two species. Breached pairing solutions with two Fermi surfaces
are also not observed. However, in the BEC region with $\bar\mu<m$, 
stable gapless 
modes are possible. The quasi particle can become gapless
for $\eta(\equiv 1/k_fa)  > 1.9$ and even for larger value of $\eta$, the 
quasi anti-fermions can also become gapless. Such gapless modes will
be relevant for
the transport coefficients of the fermionic system.

We have not calculated here the Meissner masses, or the number susceptibility
to discuss the stability of different phases by ruling out regions in the
parameter space of gap and chemical potential difference. Instead, we have solved the gap equation and the number 
density equation self consistently and have compared the
thermodynamic potentials.  In certain regions of the chemical potential
difference and the coupling, we have multiple solutions for the gap equation.
In such cases, we have taken the solution which has the lowest
thermodynamic potential, ensuring that it corresponds to a minimum.
In the deep BEC region, the phase 
transition from BCS to gapless phase is a second order phase transition with the order parameter 
decreasing continuously, while the transition from the
gapless phase to the normal matter phase is a first order transition as the
difference in the densities of the two condensing species is increased.

The results obtained here are of course limited by the choice of the ansatz.
We have not considered here other nonuniform ansatz leading to
the LOFF phase or the crystalline phase \cite{nardulli,amhmloff,ren,rishi}. 
The results obtained here might nevertheless
be regarded as a reference solution with which other numerical or analytical results
obtained from more involved ansatz for the ground state may be compared.
Though it is not apriori obvious, the results, obtained through the simple variational
ansatz for the ground state, regarding the phase structure for 
this purely fermionic theory turn out to be similar to those of 
a Bose fermi model of Ref.\cite{andreas} treated within a mean
field approximation.

We have also considered the effects of the fluctuations by treating the 
condensate field as a dynamical bosonic field in a model with quartic 
self interactions of the boson field. The BCS ansatz was modified to include the quanta of the fluctuating field along with the usual fermion pairs.
Such an ansatz gave rise to a superfluid gap equation that includes 
the effect of condensate fluctuations. In the evaluation of the 
superconducting gap
the scalar field mass gap was also calculated self consistently. This leads
to a decrease of the critical temperature in the BEC regime. We also 
observed that the superfluid transition could be first order for larger
quartic coupling with the effect of the condensate fluctuations becoming larger. 
The present ansatz for the ground state leads to the result arising from a
summation of an infinite series of bubble diagrams for the scalar field.
However, this 
does not include the effect of 'sunset' type diagram. Inclusion of 
such diagrams has been successfully done recently within a CJT formulation 
in Ref.\cite{deng}.

 We can generalize this toy model to a more realistic
model like Nambu Jona-Lasinio model, to describe the possibility of 
relativistic BCS-BEC crossover in quark matter phase diagram. Further, 
the effect of charge neutrality conditions, as appropriate for matter 
in the core of neutron stars, on relativistic BCS-BEC crossover will 
be important. Some of these calculations are in progress and
will be reported elsewhere.

\acknowledgments

HM would like to thank the organizers of the meeting on 
``Interface of QGP and cold atoms" at ECT$^*$, Trento and 
acknowledges discussions with D.H. Rischke, Y. Nishida, H. Abuki.
AM would like to thank Frankfurt Institute for 
Advanced Studies (FIAS) for warm hospitality,
where the present work was initiated and Alexander von Humboldt Foundation, 
Germany for financial support during the visit. 
HM would also like to thank Institut fuer Theoretische Physik, University
of Frankfurt for hospitality and Alexander von Humboldt Foundation for support.
AM would like to acknowledege financial support from Department of Science
and Technology, Government of India (project
no. SR/S2/HEP-21/2006).

\def \ltg{R.P. Feynman, Nucl. Phys. B 188, 479 (1981); 
K.G. Wilson, Phys. Rev. \zbf  D10, 2445 (1974); J.B. Kogut,
Rev. Mod. Phys. \zbf  51, 659 (1979); ibid  \zbf 55, 775 (1983);
M. Creutz, Phys. Rev. Lett. 45, 313 (1980); ibid Phys. Rev. D21, 2308
(1980); T. Celik, J. Engels and H. Satz, Phys. Lett. B129, 323 (1983)}

\def\loff{A.I. Larkin and Yu.N. Ovchinnikov, Sov. Phys. JETP{\bf 20} (1965);
P. Fulde and R.A. Ferrel, Phys Rev. {\bf A135}, 550, 1964.}
\def\takada{S. Takada and T. Izuyama, Prog. theor. Phys. {\bf 41}, 635 (1969).}
\def\ren{I. Giannakis, H. Ren,{\PLB{611}{137}{2005}};
{\em ibid}{\NPB{723}{255}{2005}}.}
\def\berges {J. Berges, K. Rajagopal, {\NPB{538}{215}{1999}}.}
\def \svz {M.A. Shifman, A.I. Vainshtein and V.I. Zakharov,
Nucl. Phys. B147, 385, 448 and 519 (1979);
R.A. Bertlmann, Acta Physica Austriaca 53, 305 (1981)}

\def \spmbst {S.P. Misra, Phys. Rev. D35, 2607 (1987)}

\def \hmgrnv { H. Mishra, S.P. Misra and A. Mishra,
Int. J. Mod. Phys. A3, 2331 (1988)}

\def \snss {A. Mishra, H. Mishra, S.P. Misra
and S.N. Nayak, Phys. Lett 251B, 541 (1990)}

\def \amqcd { A. Mishra, H. Mishra, S.P. Misra and S.N. Nayak,
Pramana (J. of Phys.) 37, 59 (1991). }
\def\qcdtb{A. Mishra, H. Mishra, S.P. Misra 
and S.N. Nayak, Z.  Phys. C 57, 233 (1993); A. Mishra, H. Mishra
and S.P. Misra, Z. Phys. C 58, 405 (1993)}

\def \spmtlk {S.P. Misra, Talk on {\it `Phase transitions in quantum field
theory'} in the Symposium on Statistical Mechanics and Quantum field theory, 
Calcutta, January, 1992, hep-ph/9212287}

\def \hmnj {H. Mishra and S.P. Misra, 
{\PRD{48}{5376}{1993}.}}

\def \hmqcd {A. Mishra, H. Mishra, V. Sheel, S.P. Misra and P.K. Panda,
hep-ph/9404255 (1994)}

\def \amcrl {A. Mishra, H. Mishra and S.P. Misra, Z. Phys. C 57, 241 (1993)}

\def \higgs { S.P. Misra, in {\it Phenomenology in Standard Model and Beyond}, 
Proceedings of the Workshop on High Energy Physics Phenomenology, Bombay,
edited by D.P. Roy and P. Roy (World Scientific, Singapore, 1989), p.346;
A. Mishra, H. Mishra, S.P. Misra and S.N. Nayak, Phys. Rev. D44, 110 (1991)}

\def \nmtr {A. Mishra, 
H. Mishra and S.P. Misra, Int. J. Mod. Phys. A5, 3391 (1990); H. Mishra,
 S.P. Misra, P.K. Panda and B.K. Parida, Int. J. Mod. Phys. E 1, 405, (1992);
 {\it ibid}, E 2, 547 (1993); A. Mishra, P.K. Panda, S. Schrum, J. Reinhardt
and W. Greiner, to appear in Phys. Rev. C}

\def \dtrn {P.K. Panda, R. Sahu and S.P. Misra, 
Phys. Rev C45, 2079 (1992)}

\def \qcd {G. K. Savvidy, Phys. Lett. 71B, 133 (1977);
S. G. Matinyan and G. K. Savvidy, Nucl. Phys. B134, 539 (1978); N. K. Nielsen
and P. Olesen, Nucl.  Phys. B144, 376 (1978); T. H. Hansson, K. Johnson,
C. Peterson Phys. Rev. D26, 2069 (1982)}

\def \cornwal {J.M. Cornwall, Phys. Rev. D26, 1453 (1982)}
\def\aichlin {F. Gastineau, R. Nebauer and J. Aichelin,
{\PRC{65}{045204}{2002}}.}

\def \mndglv {J. E. Mandula and M. Ogilvie, Phys. Lett. 185B, 127 (1987)}

\def \schwinger {J. Schwinger, Phys. Rev. 125, 1043 (1962); ibid,
127, 324 (1962)}

\def \schutte {D. Schutte, Phys. Rev. D31, 810 (1985)}

\def \amspm {A. Mishra and S.P. Misra, 
{\ZPC{58}{325}{1993}}.}

\def \gft{ For gauge fields in general, see e.g. E.S. Abers and 
B.W. Lee, Phys. Rep. 9C, 1 (1973)}

\def \gribov {V.N. Gribov, Nucl. Phys. B139, 1 (1978)}

\def \spm78 {S.P. Misra, Phys. Rev. D18, 1661 (1978); {\it ibid}
D18, 1673 (1978)} 

\def \lopr {A. Le Youanc, L.  Oliver, S. Ono, O. Pene and J.C. Raynal, 
Phys. Rev. Lett. 54, 506 (1985)}

\def \spphi {S.P. Misra and S. Panda, Pramana (J. Phys.) 27, 523 (1986);
S.P. Misra, {\it Proceedings of the Second Asia-Pacific Physics Conference},
edited by S. Chandrasekhar (World Scientific, 1987) p. 369}

\def\spmdif {S.P. Misra and L. Maharana, Phys. Rev. D18, 4103 (1978); 
    S.P. Misra, A.R. Panda and B.K. Parida, Phys. Rev. Lett. 45, 322 (1980);
    S.P. Misra, A.R. Panda and B.K. Parida, Phys. Rev. D22, 1574 (1980)}

\def \spmvdm {S.P. Misra and L. Maharana, Phys. Rev. D18, 4018 (1978);
     S.P. Misra, L. Maharana and A.R. Panda, Phys. Rev. D22, 2744 (1980);
     L. Maharana,  S.P. Misra and A.R. Panda, Phys. Rev. D26, 1175 (1982)}

\def\spmthr {K. Biswal and S.P. Misra, Phys. Rev. D26, 3020 (1982);
               S.P. Misra, Phys. Rev. D28, 1169 (1983)}

\def \spmstr { S.P. Misra, Phys. Rev. D21, 1231 (1980)} 

\def \spmjet {S.P. Misra, A.R. Panda and B.K. Parida, Phys. Rev Lett. 
45, 322 (1980); S.P. Misra and A.R. Panda, Phys. Rev. D21, 3094 (1980);
  S.P. Misra, A.R. Panda and B.K. Parida, Phys. Rev. D23, 742 (1981);
  {\it ibid} D25, 2925 (1982)}

\def \arpftm {L. Maharana, A. Nath and A.R. Panda, Mod. Phys. Lett. 7, 
2275 (1992)}

\def \van {R. Van Royen and V.F. Weisskopf, Nuov. Cim. 51A, 617 (1965)}

\def \rchpi {S.R. Amendolia {\it et al}, Nucl. Phys. B277, 168 (1986)}

\def \chrl{ Y. Nambu, {\PRL{4}{380}{1960}};
A. Amer, A. Le Yaouanc, L. Oliver, O. Pene and
J.C. Raynal,{\PRL{50}{87}{1983a}};{\em ibid}
{\PRD{28}{1530}{1983}};
M.G. Mitchard, A.C. Davis and A.J.
MAacfarlane, {\NPB{325}{470}{1989}};
B. Haeri and M.B. Haeri,{\PRD{43}{3732}{1991}};
V. Bernard,{\PRD{34}{1604}{1986}};
 S. Schramm and
W. Greiner, Int. J. Mod. Phys. \zbf E1, 73 (1992), 
J.R. Finger and J.E. Mandula, Nucl. Phys. \zbf B199, 168 (1982),
S.L. Adler and A.C. Davis, Nucl. Phys.\zbf  B244, 469 (1984),
S.P. Klevensky, Rev. Mod. Phys.\zbf  64, 649 (1992).}

\def \spmijp { S.P. Misra, Ind. J. Phys. 61B, 287 (1987)}

\def \feynman {R.P. Feynman and A.R. Hibbs, {\it Quantum mechanics and
path integrals}, McGraw Hill, New York (1965)}

\def \glstn{ J. Goldstone, Nuov. Cim. \zbf 19, 154 (1961);
J. Goldstone, A. Salam and S. Weinberg, Phys. Rev. \zbf  127,
965 (1962)}

\def \anderson {P.W. Anderson, Phys. Rev. \zbf {110}, 827 (1958)}

\def \nambu{ Y. Nambu, Phys. Rev. Lett. \zbf 4, 380 (1960)}

\def\donogh {J.F. Donoghue, E. Golowich and B.R. Holstein, {\it Dynamics
of the Standard Model}, Cambridge University Press (1992)}

\def\satz {T. Matsui and H. Satz, Phys. Lett. B178, 416 (1986)}

\def\cps {C. P. Singh, Phys. Rep. 236, 149 (1993)}

\def\prliop {A. Mishra, H. Mishra, S.P. Misra, P.K. Panda and Varun
Sheel, Int. J. of Mod. Phys. E 5, 93 (1996)}

\def\hmcor {V. Sheel, H. Mishra and J.C. Parikh, Phys. Lett. B382, 173
(1996); {\it biid}, to appear in Int. J. of Mod. Phys. E}
\def\cort { V. Sheel, H. Mishra and J.C. Parikh, Phys. ReV D59,034501 (1999);
{\it ibid}Prog. Theor. Phys. Suppl.,129,137, (1997).}

\def\surcor {E.V. Shuryak, Rev. Mod. Phys. 65, 1 (1993)} 

\def\stevenson {A.C. Mattingly and P.M. Stevenson, Phys. Rev. Lett. 69,
1320 (1992); Phys. Rev. D 49, 437 (1994)}

\def\mac {M. G. Mitchard, A. C. Davis and A. J. Macfarlane,
 Nucl. Phys. B 325, 470 (1989)} 
\def\tfd
 {H.~Umezawa, H.~Matsumoto and M.~Tachiki {\it Thermofield dynamics
and condensed states} (North Holland, Amsterdam, 1982) ;
P.A.~Henning, Phys.~Rep.253, 235 (1995).}
\def\amph4{Amruta Mishra and Hiranmaya Mishra,
{\JPG{23}{143}{1997}}.}

\def \neglecor{M.-C. Chu, J. M. Grandy, S. Huang and 
J. W. Negele, Phys. Rev. D48, 3340 (1993);
ibid, Phys. Rev. D49, 6039 (1994)}

\def\revdata {Particle Data Group, Phys. Rev. D 50, 1173 (1994)}

\def\sinp {S.P. Misra, Indian J. Phys., {\bf 70A}, 355 (1996)}
\def\hmparikh{H. Mishra and J.C. Parikh, {\NPA{679}{597}{2001}.}}
\def\krisch {M. Alford and K. Rajagopal, JHEP 0206,031,(2002)}
\def\reddy {A.W. Steiner, S. Reddy and M. Prakash,
{\PRD{66}{094007}{2002}.}}
\def\hmam {Amruta Mishra and Hiranmaya Mishra,
{\PRD{69}{014014}{2004}.}}
\def\hmampp {Amruta Mishra and Hiranmaya Mishra,
in preparation.}
\def\bryman {D.A. Bryman, P. Deppomier and C. Le Roy, Phys. Rep. 88,
151 (1982)}
\def\thooft {G. 't Hooft, Phys. Rev. D 14, 3432 (1976); D 18, 2199 (1978);
S. Klimt, M. Lutz, U. Vogl and W. Weise, Nucl. Phys. A 516, 429 (1990)}
\def\alkz { R. Alkofer, P. A. Amundsen and K. Langfeld, Z. Phys. C 42,
199(1989), A.C. Davis and A.M. Matheson, Nucl. Phys. B246, 203 (1984).}
\def\sarah {T.M. Schwartz, S.P. Klevansky, G. Papp,
{\PRC{60}{055205}{1999}}.}
\def\wil{M. Alford, K.Rajagopal, F. Wilczek, {\PLB{422}{247}{1998}};
{\it{ibid}}{\NPB{537}{443}{1999}}.}
\def\sursc{R.Rapp, T.Schaefer, E. Shuryak and M. Velkovsky,
{\PRL{81}{53}{1998}};{\it ibid}{\AP{280}{35}{2000}}.}
\def\pisarski{
D. Bailin and A. Love, {\PR{107}{325}{1984}},
D. Son, {\PRD{59}{094019}{1999}}; 
T. Schaefer and F. Wilczek, {\PRD{60}{114033}{1999}};
D. Rischke and R. Pisarski, {\PRD{61}{051501}{2000}}, 
D. K. Hong, V. A. Miransky, 
I. A. Shovkovy, L.C. Wiejewardhana, {\PRD{61}{056001}{2000}}
.}
\def\leblac {M. Le Bellac, {\it Thermal Field Theory}(Cambridge, Cambridge University
Press, 1996).}
\def\bcs{A.L. Fetter and J.D. Walecka, {\it Quantum Theory of Many
particle Systems} (McGraw-Hill, New York, 1971).}
\def\alexander{Aleksander Kocic, Phys. Rev. D33, 1785,(1986).}
\def\bubmix{F. Neumann, M. Buballa and M. Oertel,
{\NPA{714}{481}{2003}.}}
\def\kunihiro{M. Kitazawa, T. Koide, T. Kunihiro, Y. Nemeto,
{\PTP{108}{929}{2002}.}}
\def\igor{Igor Shovkovy, Mei Huang, {\PLB{564}{205}{2003}}.}
\def\prasanth{P. Jaikumar and M. Prakash,{\PLB{516}{345}{2001}}.}
\def\igorr{Mei Huang, Igor Shovkovy, {\NPA{729}{835}{2003}}.}
\def\abrikosov{A.A. Abrikosov, L.P. Gorkov, Zh. Eskp. Teor.39, 1781,
1960}
\def\krischprl{M.G. Alford, J. Berges and K. Rajagopal,
 {\PRL{84}{598}{2000}.}}
\def\hatmampp{A. Mishra and H.Mishra, in preparation}
\def\blaschke{D. Blaschke, M.K. Volkov and V.L. Yudichev,
{\EPJA{17}{103}{2003}}.}
\def\mei{M. Huang, P. Zhuang, W. Chao,
{\PRD{65}{076012}{2002}}}
\def\bubnp{M. Buballa, M. Oertel,
{\NPA{703}{770}{2002}}.}
\def\sarma{G. Sarma, J. Phys. Chem. Solids 24,1029 (1963).}
\def\ebert {D. Ebert, H. Reinhardt and M.K. Volkov,
Prog. Part. Nucl. Phys.{\bf 33},1, 1994.}
\def\rehberg{ P. Rehberg, S.P. Klevansky and J. Huefner,
{\PRC{53}{410}{1996}.}}
\def\lutz{M. Lutz, S. Klimt, W. Weise,{\NPA{542}{521}{1992}.}}
\def\rapid{B. Deb, A.Mishra, H. Mishra and P. Panigrahi,
Phys. Rev. A {\bf 70},011604(R), 2004.}
\def\kriscfl{M. Alford, C. Kouvaris, K. Rajagopal, Phys. Rev. Lett.
{\bf 92} 222001 (2004), arXiv:hep-ph/0406137.}
\def\shovris{S.B. Ruester, I.A. Shovkovy and D.H. Rischke,
arXiv:hep-ph/0405170.}
\def\krisaug{K. Fukushima, C. Kouvaris and K. Rajagopal, arxiv:hep-ph/0408322}.
\def\wilczek{W.V. Liu and F. Wilczek,{\PRL{90}{047002}{2003}},E. Gubankova,
W.V. Liu and F. Wilczek, {\PRL{91}{032001}{2003}.}}
\def\review{For reviews see
M.G. Alford, A. Schmitt, K. Rajagopal and T. Schaefer, arXiv:0709.4635
 K. Rajagopal and F. Wilczek,
arXiv:hep-ph/0011333; D.K. Hong, Acta Phys. Polon. B32,1253 (2001);
M.G. Alford, Ann. Rev. Nucl. Part. Sci 51, 131 (2001); G. Nardulli,
Riv. Nuovo Cim. 25N3, 1 (2002); S. Reddy, Acta Phys Polon.B33, 4101(2002);
T. Schaefer arXiv:hep-ph/0304281; D.H. Rischke, Prog. Part. Nucl. Phys. 52,
197 (2004); H.C. Ren, arXiv:hep-ph/0404074; M. Huang, arXiv: hep-ph/0409167;
I. Shovkovy, arXiv:nucl-th/0410191.}
\def\kunihiroo{ M. Kitazawa, T. Koide, T. Kunihiro and Y. Nemoto,
{\PRD{65}{091504}{2002}}, D.N. Voskresensky, arXiv:nucl-th/0306077.}
\def\rupak{S.Reddy and G. Rupak, arXiv:nucl-th/0405054}
\def\ida{K. Iida and G. Baym,{\PRD{63}{074018}{2001}},
Erratum-ibid{\PRD{66}{059903}{2002}}; K. Iida, T. Matsuura, M. Tachhibana 
and T. Hatsuda, {\PRL{93}{132001}{2004}}; ibid,{arXiv:hep-ph/0411356}}
\def\chromo{Mei Huang and Igor Shovkovy,{\PRD{70}{051501}{2004}};
 {\em ibid}, {\PRD{70}{094030}{2004}}}
\def\steiner{A.W. Steiner, {\PRD{72}{054024}{2005}.}}
\def\andreaskris{K. Rjagopal and A. Schimitt{\PRD{73}{045003}{2006}.}}
\def\andreas{J.Deng, A. Schmitt and Q. Wang {\PRD{76}{034013}{2007}.}}
\def\deng{J.Deng, J. Wang and Q. Wang arXiv:0803.4360.}
\def\krisandreas{K. Rajagopal and A. Schmitt, {\PRD{73}{045003}{2006}.}}
\def\abuki{H. Abuki,{\NPA{791}{117}{2007}}.}
\def\abukib{Y. Nishida and H. Abuki,{\PRD{72}{096004}{2005}.}}
\def\brauner{Tomas Brauner, arXiv:0803.2422[hep-ph].}
\def\nardulli{M. Mannarelli, G. Nardulli and M. Ruggieri, {\PRA{74}{033606}{2006}}.}
\def\rischke{M. Kitazawa, D. Rischke and I.A. Shovkovy, arXiv:0709.2235 hep-ph.}
\def\amhm5{A. Mishra and H. Mishra, {\PRD{71}{074023}{2005}.}}
\def\leupold{K. Schertler, S. Leupold and J. Schaffner-Bielich,
{\PRC{60}{025801}(1999).}}
\def\bubrep{Michael Buballa, Phys. Rep.{\bf 407},205, 2005.}
\def\hatkun{T. Hatsuda and T. Kunihiro, Phys. Rep.{\bf 247},221, 1994.}
\def\hatsuda{H. Abuki, T. Hatsuda, K. Itakura, {\PRD{65}{074014}{2002}}.}
\def\lkw{ M. Lutz, S. Klimt and W. Weise, Nucl Phys. {\bf A542}, 521, 1992.}
\def\ruester{S.B. Ruester, V.Werth, M. Buballa, I. Shovkovy, D.H. Rischke,
arXiv:nucl-th/0602018; S.B. Ruester, I. Shovkovy, D.H. Rischke,
{\NPA{743}{127}{2004}.}}
\def\hmparikh{H. Mishra and J.C. Parikh, {\NPA{679}{597}{2001}.}}
\def\amhma{Amruta Mishra and Hiranmaya Mishra,
{\PRD{69}{014014}{2004}.}}
\def\amhmb{A. Mishra and H. Mishra, {\PRD{71}{074023}{2005}.}}
\def\amhmc{A. Mishra and H. Mishra, {\PRD{74}{054024}{2006}.}}
\def\caldas{H. Caldas, arXiv:cond-mat/0605005}
\def\amhmloff{A. Mishra and H. Mishra, arXiv:cond-mat/0611058.}
\def\randeria{C.A.R. Sa de Melo, M. Randeria and J.R. Engelbrecht,
{\PRL{71}{3202}{1993}}.}
\def\carlsonreddy{S.Y. Chang, J. Carlson, V.R. Pandharipande and K.E. Schmidt,
{\PRA{70}{043602}{2004}}; J. Carlson and S. Reddy, {\PRL{95}{060401}{2005}}.}
\def\nishidason{Y. Nishida and D.T. Son, {\PRL{97}{050403}{2006}},
G. Rupak, T. Schafer and A. Kryjevski, {\PRA{75}{023606}{2007}}.}
\def\zhuang{G.F. Sun, L.He, P. Zhuang, {\PRD{75}{096004}{2007}}, 
L. He, P. Zhuang {\PRD{76}{056003}{2007}}.}
\def\zhuangb{L.He, P. Zhuang, {\PRD{75}{096003}{2007}}.}
\def\chennakano{J.W. Chen and E. Nakano, {\PRA{75}{043620}{2007}}.}
\def\rishi{K.Rajagopal and R. Sharma, {\PRD{74}{094019}{2006}}}
\def\cameliapi{G. Amelino-Camelia and S.Y. Pi,{\PRD{47}{2356}{1993}.}}
\def\rislena{D. Rischke and J. Lenaghan,{\JPG{26}{431}{2000}.}}
\def\risph{I. Giannakis, D.Hou and H.Ren and 
D. Rischke,{\PRL{93}{232301}{2004}}.}
\def\pirner{A.H. Rezaean and H.J. Pirner, {\NPA{779}{197}{2006}}}

\end{document}